\documentclass[a4paper,onecolumn,11pt,unpublished]{quantumarticle}
\pdfoutput=1
\usepackage[utf8]{inputenc}
\usepackage[english]{babel}
\usepackage[T1]{fontenc}

\usepackage[usenames,dvipsnames]{color}
\usepackage[numbers]{natbib}
\usepackage[hyperfootnotes=false]{hyperref}
\hypersetup{
	colorlinks,
	citecolor=quantumviolet,
	linkcolor=quantumviolet,
	urlcolor=quantumviolet}
\usepackage{tikz}
\usepackage{lipsum}
\usepackage[linesnumbered,ruled,vlined]{algorithm2e}
\usepackage[noend]{algpseudocode}
\usepackage{mathtools}
\usepackage{graphicx}
\usepackage{amssymb}
\usepackage{amsmath}
\usepackage{amsthm}
\usepackage{amsfonts}
\usepackage{color}
\usepackage{xcolor}
\usepackage{braket}
\usepackage{soul}
\usepackage{multirow}
\usepackage{subfigure}
\usepackage{booktabs}
\usepackage{hyperref}
\usepackage{threeparttable}
\usepackage{epstopdf}

\SetKwInput{KwInput}{Input}                
\SetKwInput{KwOutput}{Output}              

\definecolor{CCTLABgreen}{RGB}{0,128,0}
\usepackage{array}
\newcolumntype{P}[1]{>{\centering\arraybackslash}p{#1}}

\usepackage[printonlyused,withpage]{acronym}

\newacro{DNOT}[D-CNOT]{distributed controlled NOT}
\newacro{QZ}[QZ]{quantum Zeno}
\newacro{CQZ}[CQZ]{chained QZ}
\newacro{IFM}[IFM]{interaction-free measurement}
\newacro{BEC}[BEC]{binary erasure channel}
\newacro{QEC}[QEC]{quantum erasure channel}
\newacro{DQZ}[DQZ]{dual QZ}
\newacro{MQZ}[MQZ]{modified QZ}
\newacro{DMQZ}[D-MQZ]{dual MQZ}
\newacro{DDNOT}[DD-CNOT]{dual D-CNOT}
\newacro{DCQZ}[DCQZ]{dual CQZ}
\newacro{AO}[AO]{absorptive object}

\usepackage{bm}
\usepackage{graphicx}

\DeclareMathAlphabet{\mathsfbr}{OT1}{cmss}{m}{n}
\SetMathAlphabet{\mathsfbr}{bold}{OT1}{cmss}{bx}{n}
\DeclareRobustCommand{\msf}[1]{%
  \ifcat\noexpand#1\relax\msfgreek{#1}\else\mathsfbr{#1}\fi
}

\makeatletter
\newcommand{\msfgreek}[1]{\csname s\expandafter\@gobble\string#1\endcsname}
\makeatother

\DeclareFontEncoding{LGR}{}{} 
\DeclareSymbolFont{sfgreek}{LGR}{cmss}{m}{n}
\SetSymbolFont{sfgreek}{bold}{LGR}{cmss}{bx}{n}
\DeclareMathSymbol{\salpha}{\mathord}{sfgreek}{`a}
\DeclareMathSymbol{\sbeta}{\mathord}{sfgreek}{`b}
\DeclareMathSymbol{\sgamma}{\mathord}{sfgreek}{`g}
\DeclareMathSymbol{\sdelta}{\mathord}{sfgreek}{`d}
\DeclareMathSymbol{\sepsilon}{\mathord}{sfgreek}{`e}
\DeclareMathSymbol{\szeta}{\mathord}{sfgreek}{`z}
\DeclareMathSymbol{\seta}{\mathord}{sfgreek}{`h}
\DeclareMathSymbol{\stheta}{\mathord}{sfgreek}{`j}
\DeclareMathSymbol{\siota}{\mathord}{sfgreek}{`i}
\DeclareMathSymbol{\skappa}{\mathord}{sfgreek}{`k}
\DeclareMathSymbol{\slambda}{\mathord}{sfgreek}{`l}
\DeclareMathSymbol{\smu}{\mathord}{sfgreek}{`m}
\DeclareMathSymbol{\snu}{\mathord}{sfgreek}{`n}
\DeclareMathSymbol{\sxi}{\mathord}{sfgreek}{`x}
\DeclareMathSymbol{\somicron}{\mathord}{sfgreek}{`o}
\DeclareMathSymbol{\spi}{\mathord}{sfgreek}{`p}
\DeclareMathSymbol{\srho}{\mathord}{sfgreek}{`r}
\DeclareMathSymbol{\ssigma}{\mathord}{sfgreek}{`s}
\DeclareMathSymbol{\stau}{\mathord}{sfgreek}{`t}
\DeclareMathSymbol{\supsilon}{\mathord}{sfgreek}{`u}
\DeclareMathSymbol{\sphi}{\mathord}{sfgreek}{`f}
\DeclareMathSymbol{\schi}{\mathord}{sfgreek}{`q}
\DeclareMathSymbol{\spsi}{\mathord}{sfgreek}{`y}
\DeclareMathSymbol{\somega}{\mathord}{sfgreek}{`w}

\DeclareMathSymbol{\svarsigma}{\mathord}{sfgreek}{`c}

\DeclareMathSymbol{\sGamma}{\mathalpha}{sfgreek}{`G}
\DeclareMathSymbol{\sDelta}{\mathalpha}{sfgreek}{`D}
\DeclareMathSymbol{\sTheta}{\mathalpha}{sfgreek}{`J}
\DeclareMathSymbol{\sLambda}{\mathalpha}{sfgreek}{`L}
\DeclareMathSymbol{\sXi}{\mathalpha}{sfgreek}{`X}
\DeclareMathSymbol{\sPi}{\mathalpha}{sfgreek}{`P}
\DeclareMathSymbol{\sSigma}{\mathalpha}{sfgreek}{`S}
\DeclareMathSymbol{\sUpsilon}{\mathalpha}{sfgreek}{`U}
\DeclareMathSymbol{\sPhi}{\mathalpha}{sfgreek}{`F}
\DeclareMathSymbol{\sPsi}{\mathalpha}{sfgreek}{`Y}
\DeclareMathSymbol{\sOmega}{\mathalpha}{sfgreek}{`W}

\DeclareRobustCommand{\mcal}[1]{%
  \ifcat\noexpand#1\relax\mathnormal{#1}\else\cal{#1}\fi
}
\DeclareRobustCommand{\BM}[1]{%
  \ifcat\noexpand#1\relax\bm{\boldUppercaseItalicGreek{#1}}\else\bm{#1}\fi
}
\makeatletter
\newcommand{\boldUppercaseItalicGreek}[1]{\csname var\expandafter\@gobble\string#1\endcsname}
\makeatother
\newcommand{\rv}[1]{\msf{#1}}

\newcommand{\M}[1]{\BM{#1}}

\DeclareMathVersion{timesmath}
\SetSymbolFont{letters}{timesmath}{OML}{ntxmi}{m}{it}
\DeclareMathVersion{timesmathbold}
\SetSymbolFont{letters}{timesmathbold}{OML}{ntxmi}{b}{it}

\newcommand{\slashslash}[1]{%
  \raisebox{#1}{%
    \scalebox{.7}{%
      \rotatebox[origin=c]{18}{$-$}%
    }%
  }%
}
\newcommand{\bslash}{%
  \mbox{%
   \vphantom{b}%
   \ooalign{\kern-.1em\smash{\slashslash{.9ex}}\hidewidth\cr$b$\cr}%
   \kern.05em
  }%
}
\newcommand{\slashed}[4]{%
  \mbox{%
   \vphantom{b}%
   \ooalign{\kern-#3\smash{\slashslash{#2}}\hidewidth\cr$#1$\cr}%
   \kern#4
  }}%
  
\begin{document}

\title{Counterfactual Full-Duplex Communication}

\author{Fakhar Zaman}
\orcid{0000-0002-3751-8933}
\author{Hyundong Shin}
\orcid{0000-0003-3364-8084}
\affiliation{Department of Electronics and Information Convergence Engineering, Kyung Hee University,\\ 1732 Deogyeong-daero, Yongin-si, Gyeonggi-do 17104, Korea}
\email{hshin@khu.ac.kr}
\thanks{(corresponding author)}
\author{Moe Z. Win}
\affiliation{Laboratory for Information and Decision Systems (LIDS), Massachusetts Institute of Technology,\\ 77 Massachusetts Avenue, Cambridge, MA 02139 USA.}
\maketitle

\begin{abstract}
This paper proposes two new full-duplex quantum communication protocols to exchange  classical or quantum information between two remote parties simultaneously  without transferring a  physical particle over the quantum channel.
The first protocol, called \emph{quantum duplex coding}, enables to exchange of a classical bit  using a preshared maximally entangled pair of qubits by means of counterfactual disentanglement.
The second protocol, called \emph{quantum telexchanging}, enables to exchange an arbitrary unknown qubit  without using preshared entanglement by means of counterfactual entanglement and disentanglement. 
We demonstrate the quantum duplex coding and quantum telexchanging by exploiting counterfactual electron-photon interaction gates and show that these quantum duplex communication protocols form full-duplex binary erasure channel and quantum erasure channels, respectively.
\end{abstract}

\section{Introduction} \label{sec 1}

A duplex communication system is to convey information between remote parties in both directions, whereas simplex communication allows sending the information in one direction only. In a half-duplex system, remote parties can transfer information in both directions but not at the same time. Full-duplex communication is the way to transfer information in both directions simultaneously. 
In classical (radio frequency) communication, the full-duplex capability is typically achieved by channelization and/or transceiver configurations \cite{ZLVH:16:IEEE}. In quantum mechanics, full-duplex communication can be achieved by means of quantum dialogue \cite{WZLH:16:QIP,M:17:QIP,ZSYL:20:SB,ZDSZSG:17:PRL} and quantum state exchange protocols \cite{LTYAL:19:PRL,LYAL:19:PRA,OW:05:arXiv,LAYZ:19:PRA} based on the laws of quantum physics such as quantum entanglement \cite{FRJKS:19:SR,HHHH:09:RMP,BLPK:05:PRL}.

Quantum dialogue \cite{WZLH:16:QIP,M:17:QIP,ZSYL:20:SB,ZDSZSG:17:PRL} provides the novel way of exchanging classical information by utilizing the preshared entanglement,
which is unattainable in classical communication. 
Although the quantum dialogue allows to achieve full-duplex communication, it requires multiple entangled pairs 
to exchange even one bit of classical information. To date, it is not possible to transfer classical information in both directions by using only one entangled pair.  This paper proposes a full-duplex protocol which allows remote parties to exchange one bit of classical information by using only one entangled pair but no physical particle is transmitted over the quantum channel.
In contrast, quantum state exchange allows remote parties to exchange arbitrary quantum states by means of \emph{preshared entanglement}, local operations and classical announcements  \cite{LTYAL:19:PRL,LYAL:19:PRA,OW:05:arXiv}.  However, the amount of information exchanged between remote parties relies on the amount of preshared entanglement. 
This paper proposes a full-duplex quantum communication protocol to exchange one qubit quantum information in both directions simultaneously without using preshared entanglement by exploiting the counterfactual quantum communication.

Counterfactual quantum communication is a revolutionary achievement in quantum informatics, which enables remote parties to transmit information under a probabilisitc model of sending a physical particle over the channel \cite{SLAZ:13:PRL,AV:19:PRA,LAZ:14:PRA,CLYCYCMPP:17:PNAS,ACP:21:NC}. At the time of successful transmission of information, no physical particle is found in the transmission channel. The counterfactuality was first introduced in quantum protocols as the counterfactual quantum computation \cite{HRBPK:06:Nature,KJHWKSJD:15:PRL,MJP:07:PRA,ZSW:20:arXiv} followed by the counterfactual quantum cryptography \cite{N:09:PRL,YLYZWCGH:12:PRA,SH:14:PRA,SW:10:PRA,LJLTTZPCCCP:12:PRL}. The basic concept is originated from the \ac{IFM} to ascertain the presence or absence of an \ac{AO} in an interferometer without physically interrogating it  \cite{VE:93:FOP,KWHZK:95:PRL}. 

The direct counterfactual quantum communication is based on the \ac{CQZ} effect where a classical bit is encoded as the presence or absence of the \ac{AO} in the interferometer \cite{SLAZ:13:PRL,AV:19:PRA,Cetal:19:NPJ}. 
In contrast, transfering the quantum information requires a quantum \ac{AO} in the superposition of presence and absence states. This idea has been first demonstrated in \cite{GCCWHZ:15:SR} to transfer the quantum information counterfactually along with a one-bit classical announcement. This was extended in \cite{LAZ:15:PRA} to transfer quantum information without transmitting any physical particle over a neither quantum nor classical channel by means of controlled disentanglement. A more effieicent protocol for counterfactual communication of quantum information was presented in \cite{SH:16:FP} exploiting the \ac{DCQZ} effect.  In recent years,  the quantum protocols have been also designed for  counterfactual entanglement distribution \cite{GCCWZ:14:OPT,CGJXC:15:OPT,CJGXC:16:JOSA}, counterfactual Bell-sate analysis \cite{ZJS:18:SR,ZHS:21:QIP}, and counterfactual cloning \cite{HZCWZ:17:PRA}.

This paper proposes two new full-duplex quantum communication protocols to exchange   information between Alice and Bob  by exploiting the inherent property of counterfactual quantum communication. The first protocol, called \emph{quantum duplex coding}, enables to exchange classical information without transferring any physical particle over a neither quantum nor classical channel. The second protocol, called \emph{quantum telexchanging}, enables to exchange an arbitrary unknown qubit  without using preshared entanglement and without transferring any physical particle over the quantum channel. The protocols are accomplished by designing \emph{nonlocal} operations to transfer classical as well as quantum information in both directions at the same time counterfactually. The key contributions of this paper are as follows:
\begin{itemize}

\item 
\emph{Quantum duplex coding}: A bit exchange protocol is proposed by developing the \ac{DNOT} operation.  The protocol enables each party to exchange a one-bit classical information in each direction simultaneously using a single preshared Bell pair of qubits by means of counterfactual disetanglement. The protocol is demonstrated by devising the nonlocal \ac{DNOT} operation for Bell-type states based on the \ac{MQZ} gate. In contract to the \ac{CQZ} gate \cite{ZJS:18:SR}, the protocol is designed by using only the blocking event (presence of the absorptive object) for full counterfactuality. It is shown that the quantum duplex coding forms a full-duplex \ac{BEC}. 

\item
\emph{Quantum telexchanging:} A quantum state exchange protocol is proposed by generalizing the idea of \ac{DNOT} operation. The protocol enables remote parties to exchange an arbitrary unknown one-qubit quantum information simultaneously without using presahred entanglement by means of counterfactual entanglement and disentanglement of the \ac{DDNOT} operation. The protocol is demonstrated by devising the \ac{DDNOT} operation for general input states based on the \ac{DMQZ} and \ac{CQZ}  gates \cite{ZJS:18:SR}. It is shown that the quantum telexchanging protocol creates a full-duplex form of the \ac{QEC}.

\end{itemize}

The remaining sections are organized as follows: Section~\ref{sec 2} briefly explain the preliminaries to design the protocols for quantum duplex coding and telexchanging. In Section~\ref{sec 3}, the quantum duplex coding protocol is proposed to transfer classical information in both directions simultaneously. In Section~\ref{sec 4}, the quantum telexchanging protocol is demonstrated by using a combination of the \ac{CQZ} and  \ac{DMQZ} gates for quantum state exchange in a counterfactual way. In Section~\ref{sec 5}, the brief comparison is drawn between the quantum duplex coding (telexchaning) and quantum superdense coding (teleportation). Finally, Section~\ref{sec 6} gives our conclusions and final remarks.

\emph{Notations}:
Random variables are displayed in sans serif, upright fonts; their realizations
in serif, italic fonts. Vectors and matrices are denoted by bold lowercase and uppercase letters, respectively.

%
%

%

\section{Counterfactual Quantum Communication}\label{sec 2}
The counterfactual quantum communication is based on the single-particle nonlocality and quantum measurement theory.  A quantum state usually collapses back to its initial state if the time between repeated measurements is short enough \cite{IHBW:90:PRA}. This QZ effect  has been demonstrated to achieve IFM where the the state of a photon acts as an unstable quantum state corresponding to the presence of the absorptive object \cite{KWHZK:95:PRL}.This section begin by introducing a brief review on the overall actions of the \ac{QZ} and \ac{CQZ} gates \cite{ZJS:18:SR,ZJS:19:SR} that are invoked to formulate the \ac{DNOT} and dual \ac{DNOT} operations to transfer classical and quantum information in both directions simultaneously. 
\begin{figure}[t!]	
  \centering
\includegraphics[width=0.65\textwidth]{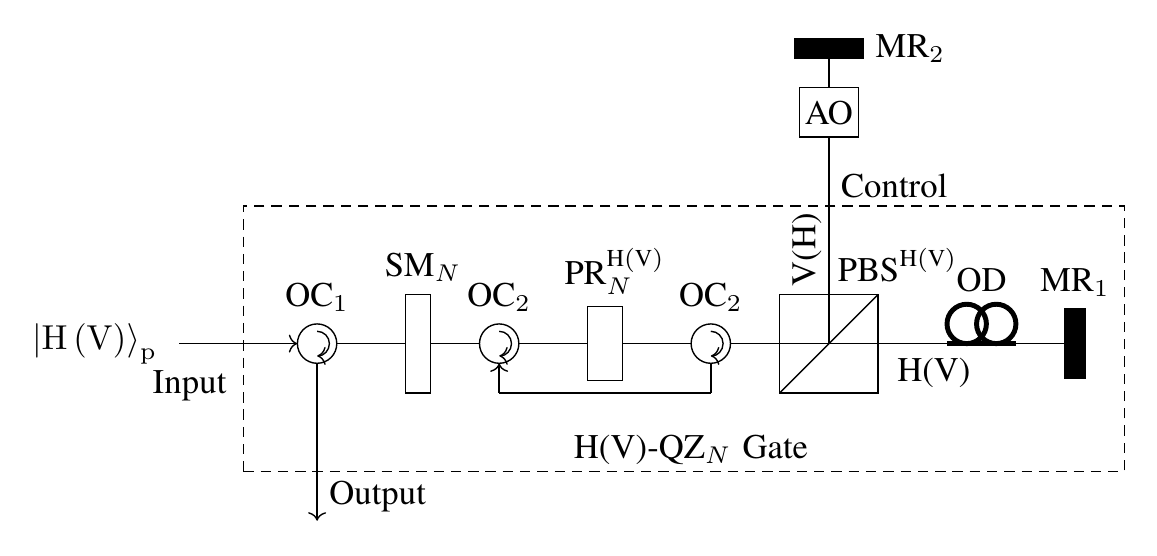}
    \caption{
     A H(V)-\ac{QZ}$_N$ gate with $N$ cycles where H (V) stands for horizontal (vertical) polarization of the photon, OC for an optical circulator, SM for a switchable mirror, PR for a polarizing rotator, PBS for a polarizing beam splitter, MR for a mirror, and AO shows the state of an absorptive object.
      }
    \label{fig:QZE}
\end{figure}

\subsection{\ac{QZ} Gates}
Fig.~\ref{fig:QZE} shows the Michelson version of the \ac{QZ} gate \cite{ZJS:19:SR} to perform \ac{IFM}. The \ac{QZ} gate is to ascertain the classical behavior of an absorptive object, i.e., to infer the absence state $\ket{0}_\mathrm{AO}$ or the presence state $\ket{1}_\mathrm{AO}$ of AO without interacting with it. The H(V)-\ac{QZ}$_N$ gate takes an H~(V) polarized photon as input. The switchable mirror SM$_N$ is initially turned off to allow passing the photon and is turned on for $N$ cycles once the photon is passed. After $N$ cycles, SM$_N$ is turned off again allowing the photon out.  The polarization rotator PR$_N^{\mathrm{H}\left(\mathrm{V}\right)}$ gives rotation to the input photon by an angle $\theta_N=\pi/\left(2N\right)$ as follows:
\begin{align}
\text{PR}_N^{\mathrm{H}\left(\mathrm{V}\right)}:
	\begin{cases}
	\ket{\mathrm{H}\left(\mathrm{V}\right)}_{\mathrm{p}}
		\rightarrow
		\cos\theta_N
		\ket{\mathrm{H}\left(\mathrm{V}\right)}_{\mathrm{p}}
		+
		\sin\theta_N
		\ket{\mathrm{V}\left(\mathrm{H}\right)}_{\mathrm{p}},\\
	\ket{\mathrm{V}\left(\mathrm{H}\right)}_{\mathrm{p}}
		\rightarrow
		\cos\theta_N
		\ket{\mathrm{V}\left(\mathrm{H}\right)}_{\mathrm{p}}
		-
		\sin\theta_N
		\ket{\mathrm{H}\left(\mathrm{V}\right)}_{\mathrm{p}}.
	\end{cases}
\end{align}
The photon state $\ket{\phi}$ after PR$_N^{\mathrm{H}\left(\mathrm{V}\right)}$ in the first cycle of the H(V)-\ac{QZ}$_N$ gate is given by
\begin{align}
\ket{\phi}=\cos\theta_N\ket{\mathrm{H}\left(\mathrm{V}\right)}_{\mathrm{p}}+\sin\theta_N\ket{\mathrm{V}\left(\mathrm{H}\right)}_{\mathrm{p}}. \label{eq:QZ pass}
\end{align}
Then, the polarizing beam splitter PBS separates the H and V components of the photon into two different optical paths: SM $\rightarrow$ MR$_1$ and SM $\rightarrow$ MR$_2$. The H~(V) component  goes towards MR$_1$ and the V~(H) component goes towards MR$_2$. The photon component in the second optical path only interacts with AO (control terminal). 

\begin{itemize}
\item 
$\text{AO}=\ket{0}_\mathrm{AO}$:
In the absence of the absorptive object, the V~(H) component of the photon is reflected by MR$_2$ and is returned back to PBS. Hence, the photon state remains unchanged. After $n \left(<N\right)$ cycles, the photon state is given by
\begin{align}	\label{eq:state:n}
\ket{\phi} 
	=
	\cos\left(n\theta_N\right)
	\ket{\mathrm{H}\left(\mathrm{V}\right)}_{\mathrm{p}}
	+
	\sin\left(n\theta_N\right)
	\ket{\mathrm{V}\left(\mathrm{H}\right)}_{\mathrm{p}}.
\end{align} 
The photon will end up in the state $\ket{\mathrm{V}\left(\mathrm{H}\right)}_{\mathrm{p}}$ with certainty by $\pi/2$ rotation after $N$ cycles.
\begin{table*}[t!]
\centering
\setlength{\tabcolsep}{8pt}
\renewcommand{\arraystretch}{1.5}
\caption{H(V)-\ac{QZ}$_N$ and H(V)-\ac{CQZ}$_{M,N}$ gates.}\label{tab:I-QZE}
\vspace{0.2cm}
\begin{threeparttable}
\begin{tabular}{P{1.2cm} |  P{1.2cm} | P{1.2cm} | P{1.8cm} | P{1.0cm} | P{1.2cm} | P{1.8cm} | P{1.0cm}}
\hline
\multirow{2}{*}{Input} & \multirow{2}{*}{Control} &  \multicolumn{3}{c|}{\ac{QZ} Gate} & \multicolumn{3}{c}{\ac{CQZ} Gate}  \\
 \cline{3-8}
 & &  Output & Probability & CF & Output & Probability & CF\\
\hline
\hline 
\multirow{2}{*}{$\ket{\mathrm{H}\left(\mathrm{V}\right)}_{\mathrm{p}}$} & $\ket{0}_{\mathrm{AO}}$ &  $\ket{\mathrm{V}\left(\mathrm{H}\right)}_{\mathrm{p}}$ & 1 & No & $\ket{\mathrm{H}\left(\mathrm{V}\right)}_{\mathrm{p}}$ & $\lambda_0$ & Yes\\
& $\ket{1}_{\mathrm{AO}}$ &  $\ket{\mathrm{H}\left(\mathrm{V}\right)}_{\mathrm{p}}$ &  $\cos^{2N}\theta_N$ & Yes & $\ket{\mathrm{V}\left(\mathrm{H}\right)}_{\mathrm{p}}$ & $\lambda_1$ & Yes\\[0.5ex]
\hline
\end{tabular}
\begin{tablenotes}

\item[Note)]
CF: Counterfactuality

\end{tablenotes}
\end{threeparttable}
\end{table*}

\item 
$\text{AO}=\ket{1}_\mathrm{AO}$:
In the presence of the absorptive object, the V~(H) component is absorbed by AO if it is found in the control terminal. In each cycle, the probability of this absorption event is equal to $\sin^2\theta_N$. Unless the photon is absorbed, the photon state collapses to the initial state $\ket{\mathrm{H}\left(\mathrm{V}\right)}_{\mathrm{p}}$. 
After $N$ cycles, the photon is not absorbed and ends up in the state $\ket{\mathrm{H}\left(\mathrm{V}\right)}_{\mathrm{p}}$ with probability $\cos^{2N}\theta_N$ tending to one as $N\rightarrow\infty$.
\end{itemize}

Table~\ref{tab:I-QZE} shows the overall action of the \ac{QZ} gate. Note that the H(V)-\ac{QZ}$_N$ gate has the output $\ket{\mathrm{H}\left(\mathrm{V}\right)}_{\mathrm{p}}$ in the presence state $\ket{1}_\mathrm{AO}$  if the photon has not traveled over the control terminal (quantum channel). Hence, the \ac{QZ} gate is counterfactual only for this measurement outcome.

\subsection{\ac{CQZ} Gates}
\begin{figure}[t!]
 \centering{
\includegraphics[width=0.425\textwidth]{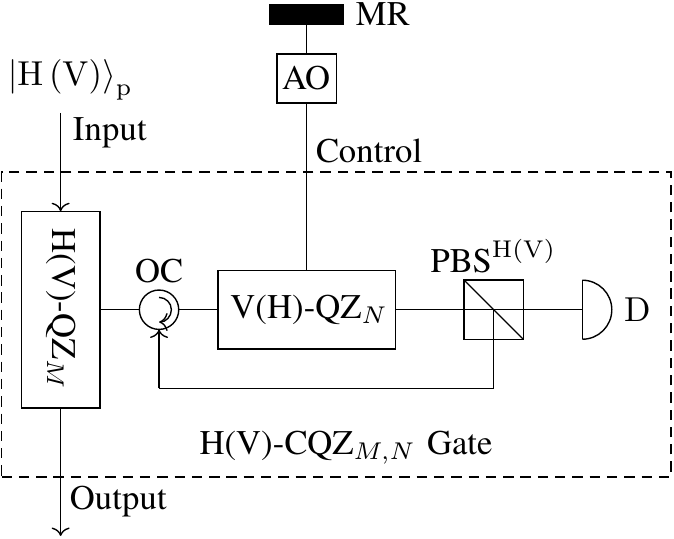}}
    \caption{
   A H(V)-\ac{CQZ}$_{M,N}$ gate with $M$ outer and $N$ inner cycles where D is a photon detector. Table~\ref{tab:I-QZE} shows the overall action of the H(V)-\ac{CQZ}$_{M,N}$ gate.    
   }
    \label{fig:MCQZ}
\end{figure}

Fig.~\ref{fig:MCQZ} shows the nested version of \ac{QZ} gates with $M$ outer and $N$ inner cycles \cite{ZJS:18:SR}. The \ac{CQZ} gate enables to ascertain the absence or presence of the absorptive object counterfactually for both the outcomes. The H(V)-\ac{CQZ}$_{M,N}$ gate also takes an H~(V) polarized photon as input. In each outer cycle, the V~(H) component of the photon enters the inner V(H)-\ac{QZ}$_N$ gate.

\begin{itemize}

\item 
$\text{AO}=\ket{0}_\mathrm{AO}$:
In the absence of the absorptive object, the inner V(H)-\ac{QZ}$_N$ gate transforms the photon state $\ket{\mathrm{V}\left(\mathrm{H}\right)}_{\mathrm{p}}$ into $\ket{\mathrm{H}\left(\mathrm{V}\right)}_{\mathrm{p}}$ after $N$ cycles.
This component ends up at the detector D after PBS. Hence, the inner \ac{QZ} gate acts as an absorptive object for the outer \ac{QZ} gate in the absence state $\ket{0}_\mathrm{AO}$, where D serves to detect the event that the photon is found in the control terminal. In each outer cycle, unless the photon is discarded, the photon state collapses back to the initial state $\ket{\mathrm{H}\left(\mathrm{V}\right)}_{\mathrm{p}}$ with probability $\cos^2 \theta_M$. After $M$ outer cycles, the photon is not discarded at the detector D and ends up in the initial state $\ket{\mathrm{H}\left(\mathrm{V}\right)}_{\mathrm{p}}$ with probability 
\begin{align}
\lambda_0
	=
	\cos^{2M} \theta_M.
\end{align}
tending to one as $M \rightarrow \infty$.

\item 
$\text{AO}=\ket{1}_\mathrm{AO}$:
In case the absorptive object is present, the V~(H) component of the photon recombines with the H~(V) component and the photon state remains unchanged for the next outer cycle, unless the photon is absorbed by AO. Hence, the inner \ac{QZ} gate acts as a mirror for the outer \ac{QZ} gate in the presence state $\ket{1}_\mathrm{AO}$. After $i \left(<M\right)$ outer cycles, unless the photon is absorbed, the photon state is given by \eqref{eq:state:n}, which is again not absorbed by AO for the next outer cycle with probability 
\begin{align}
\left[
	1-
	\sin^2\left(i\theta_M\right)
	\sin^2\theta_N
\right]^{N}.
\end{align}
Hence, unless the photon is absorbed by AO, the H(V)-\ac{CQZ}$_{M,N}$ gate transforms the input state $\ket{\mathrm{H}\left(\mathrm{V}\right)}_{\mathrm{p}}$ into $\ket{\mathrm{V}\left(\mathrm{H}\right)}_{\mathrm{p}}$ with probability 
\begin{align}	\label{eq:P1}
\lambda_1
	=
	\prod_{i=1}^M
	\left[
		1-
		\sin^2\left(i\theta_M\right)
		\sin^2\theta_N
	\right]^{N}
\end{align}
tending to one as $M, N \rightarrow \infty$.

\end{itemize}

Note that the \ac{CQZ} gate is counterfactual for both the outcomes and infers the absence or presence of the absorptive object (with probability $\lambda_0$ or $\lambda_1$) but no physical particle (photon) is found in the control terminal (see Table~\ref{tab:I-QZE}).

\subsection{Counterfactual Communication}

\begin{figure}[t!]	
  \centering
\includegraphics[width=0.45\textwidth]{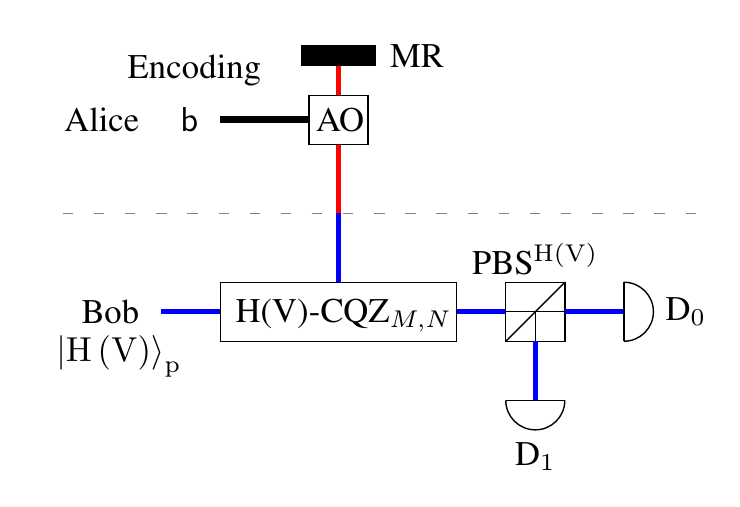}
    \caption{
H(V)-CQZ$_{M,N}$ counterfactual communication where Alice encodes her classical message $\rv{b}$ in the state $\ket{\rv{b}}_\mathrm{AO}$ of AO and Bob throws his H (V) polarized photon towards the H(V)-CQZ$_{M,N}$ gate to decode this message corresponding to the detector D$_\rv{b}$ clicks. This CQZ counterfactual communication forms a classical asymmetric BEC with the erasure probability $1-\lambda_\rv{b}$ for the message $\rv{b}$.
}
    \label{fig:CQSDC}
\end{figure}

A communication task can be achieved in a counterfactual way by using the QZ or CQZ gate where the sender (Alice) has an absorptive object and the receiver (Bob) equips the (C)QZ gate \cite{AV:19:PRA,SLAZ:13:PRL}. To transfer a classical bit $\rv{b}\in \left\{0,1\right\}$, Alice encodes this information as 
\begin{align}
\text{AO}
 =
 \ket{\rv{b}}_\mathrm{AO}.
 \end{align}
The communication with the QZ gate is counterfactual only for the one classical bit---i.e., \emph{semi-counterfactual} \cite{SLAZ:13:PRL,ZJS:19:SR}. The photon is found in the transmission channel with probability one for $\rv{b}=0$. 
To communicate both 0 and 1 without transmitting any physical particle over the transmission (quantum) channel, Bob uses the CQZ gate as shown in Fig.~\ref{fig:CQSDC}. 
Bob starts the protocol for decoding the information by throwing his H~(V) polarized photon towards the H(V)-CQZ$_{M,N}$ gate and decides that the message 0 or 1 was transmitted if it ends up in the state $\ket{\mathrm{H} \left(\mathrm{V}\right)}_{\mathrm{p}}$ or $\ket{\mathrm{V}\left(\mathrm{H}\right)}_{\mathrm{p}}$. 
That is, the CQZ receiver decides the decoded message as $\rv{b}$ if D$_\rv{b}$ clicks.  
Otherwise Bob declares that the photon is erasured (discarded or absorbed).

\begin{figure}[t!]
 \centering{
\includegraphics[width=0.6\textwidth]{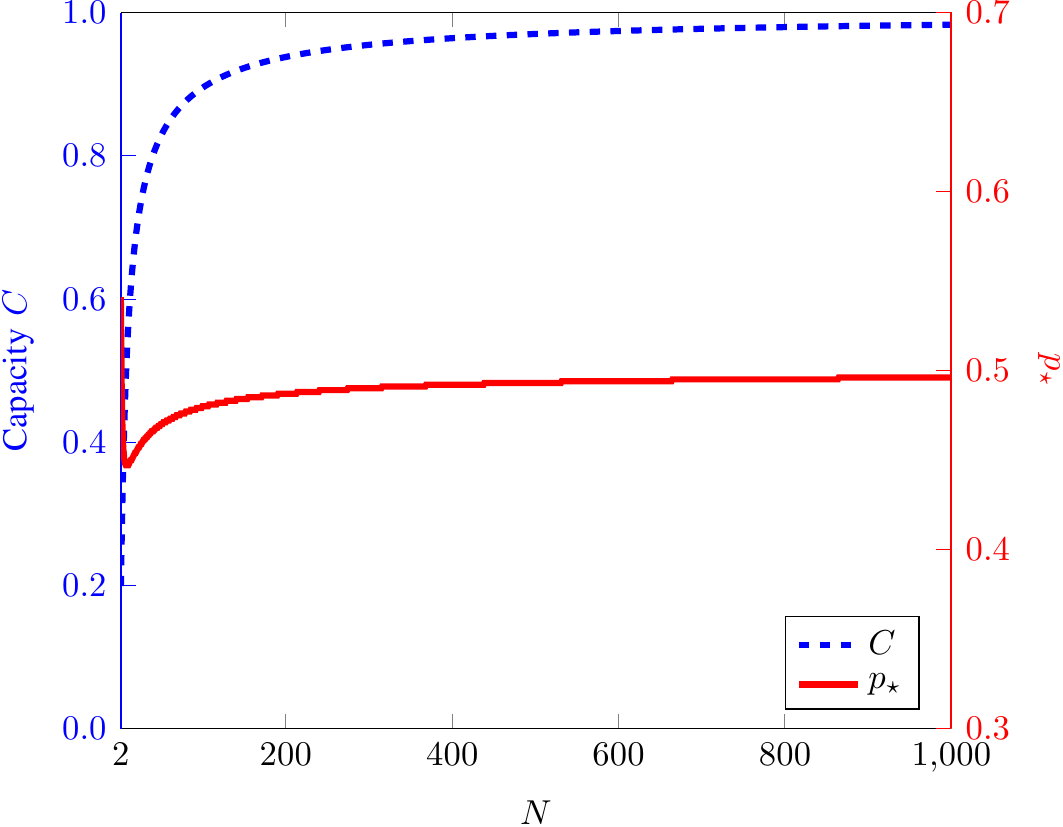}}
    \caption{
Capacity $C$ [bits/photon] and the capacity-achieving distribution $p_\star$ for the H(V)-CQZ$_{M,N}$ counterfactual communication as a function of $N$ when $M=2$. With the smallest outer cycles ($M=2$), the 80\,\% efficiency ($C=0.8$ bits/photon) is achieved at $N=81$ with $p_\star=0.466$. 
    }
    \label{fig:CAP-CQC}
\end{figure}

In case the photon is found in the transmission channel, it is either discarded by the detector in the CQZ gate (when $\rv{b}=0$ with probability $1-\lambda_0$)\footnote{In this case, Bob knows that $\rv{b}=0$ but the photon is discarded by the protocol for counterfactuality.} or absorbed by AO (when $\rv{b}=1$ with probability $1-\lambda_1$). Hence, 
this CQZ counterfactual communication forms a classical (but not symmetric) BEC \cite{CT:12:JW}. Let $p=\Pr\left[\rv{b}=1\right]$. Then, the mutual information $I \left(\mathrm{A};\mathrm{B}\right)$ between Alice ($\mathrm{A}$) and Bob ($\mathrm{B}$) is given by
\begin{align}
I\left(\mathrm{A};\mathrm{B}\right)
=
	h\left(p\right)
	-
	q 	
	h\left(
		\frac{p\left(1-\lambda_1\right)}{q}
	\right),
\end{align}
where
$h\left(p\right)=-p\log_2\left(p\right)-\left(1-p\right)\log_2\left(1-p\right)$ is the binary entropy function and
\begin{align}
q 
&=
\Pr\left[
	\text{$\rv{b}$ is erasured}
\right] 
\nonumber \\
&=
	\left(1-p\right)
	\left(1-\lambda_0\right)
	+
	p\left(1-\lambda_1\right).
\end{align}
By optimizing the message distribution $p$ such that $\bigl[\partial I\left(\mathrm{A};\mathrm{B}\right)/\partial p \bigr] \big|_{p=p_\star}=0$, we obtain the capacity $C$ in bits/photon for the CQZ counterfactual communication as follows:
\begin{align}
C
&=
	\bigl[
		I\left(\mathrm{A};\mathrm{B}\right)
	\bigr]
	\big|_{p=p_\star} 
\end{align}
taking the minimum value of 0.1515 bits/photon with $p_\star=0.606$ when $N=M=2$ and tending to 1 bit/photon with $p_\star=1/2$ as $M,N \rightarrow\infty$ (see Fig.~\ref{fig:CAP-CQC}). 

Using the dual CQZ (DCQZ) gate, the counterfactual Bell-state analysis has been proposed in \cite{ZJS:18:SR} to achieve the distinguishability task of four Bell states without transmitting physical particle over the transmission channel. In this DCQZ Bell-state analyzer, one entangled particle (electron) of the Bell pair acts as a quantum absorptive object and the other entangled particle (photon) is input to the DCQZ gate to perform the counterfactual CNOT operation. 
To improve the efficiency of quantum superdense coding, the semi-counterfactual Bell-state analyzer has been also proposed in \cite{ZJS:19:SR} using the dual QZ (DQZ) gate (instead of the DCQZ gate) with the sacrifice of full counterfactuality. This DQZ superdense coding achieves the 90\% efficiency (1.8 bits/qubit) when $N=12$.

%

%
%

\section{Quantum Duplex Coding}\label{sec 3}
In this section, we develop a full-duplex quantum protocol  to transfer classical information in both directions simultaneously and counterfactually. 
\begin{figure}[t!]
 \centering{
\includegraphics[width=0.6\textwidth]{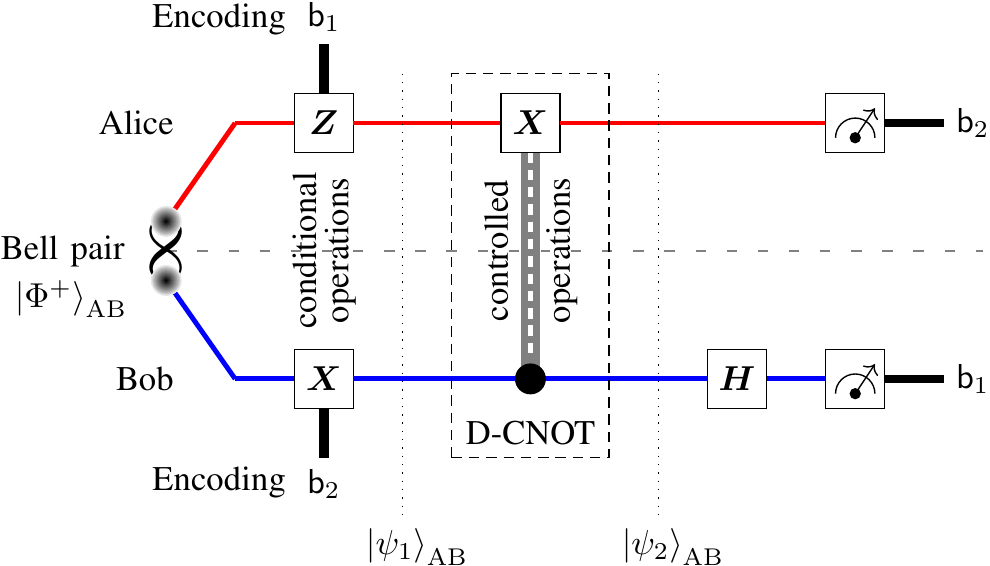}}
    \caption{
Quantum duplex coding for classical information $\rv{b}_1\rv{b}_2$. For the \ac{DNOT} operation, Alice's qubit acts as a target bit and Bob's qubit act{s} as a control bit in a counterfactual way. Here, $\M{H}$ is the Hadamard gate and $\M{Z}$ represents the Pauli $z$ operator, respectively; $\rv{b}_1$ (or $\rv{b}_2$) is the classical bit Alice (or Bob) wants to transmit to Bob (or Alice); $\ket{\psi_1}_\mathrm{AB}$ is the encoded Bell state; and $\ket{\psi_2}_\mathrm{AB}$ is the disentangled state by the \ac{DNOT} operation (viewed as counterfactual full-duplex transmission) for decoding the message.  
    }
    \label{fig:QDC}
\end{figure}
\subsection{Protocol}
Consider that Alice and Bob have a preshared maximally entangled pair (Bell state):
\begin{align}
\ket{\Phi^+}_{\mathrm{AB}}
	=
	\dfrac{1}{\sqrt{2}}
	\ket{00}_{\mathrm{AB}}
	+
	\dfrac{1}{\sqrt{2}}
	\ket{11}_{\mathrm{AB}},
\end{align}
where the subscripts A and B denote Alice and Bob, respectively. Alice and Bob encode the classical message $\rv{b}_1\rv{b}_2$ in $\ket{\psi_1}_\mathrm{AB}$ where $\rv{b}_1$ is the classical bit Alice wants to send to Bob and $\rv{b}_2$ is vice versa as follows (see Fig.~\ref{fig:QDC}):
\begin{align}
\ket{\psi_1}_\mathrm{AB}:
\begin{cases}
	00
	\rightarrow
	\left(
		\M{I}
		\otimes 
		\M{I}
	\right)
	\ket{\Phi^+}_{\mathrm{AB}}
	=
	\ket{\Phi^+}_{\mathrm{AB}}, \\
	01
	\rightarrow
	\left(
		\M{I}
		\otimes 
		\M{X}
	\right)
	\ket{\Phi^+}_{\mathrm{AB}}
	=
	\ket{\Psi^+}_{\mathrm{AB}}, \\
	10
	\rightarrow
	\left(
		\M{Z}
		\otimes 
		\M{I}
	\right)
	\ket{\Phi^+}_{\mathrm{AB}}
	=
	\ket{\Phi^-}_{\mathrm{AB}}, \\
	11
	\rightarrow
	\left(
		\M{Z}
		\otimes 
		\M{X}
	\right)
	\ket{\Phi^+}_{\mathrm{AB}}
	=
	\ket{\Psi^-}_{\mathrm{AB}},
\end{cases}
\end{align}
where $\M{I}$ is the identity operator; $\M{X}$ and $\M{Z}$ represent Pauli $x$ and $z$ operators, respectively; and 
\begin{align}
\ket{\Phi^{\pm}}_{\mathrm{AB}}
	&=
	\dfrac{1}{\sqrt{2}}
	\ket{00}_{\mathrm{AB}}
	\pm
	\dfrac{1}{\sqrt{2}}
	\ket{11}_{\mathrm{AB}},\label{eq:Phi}\\
\ket{\Psi^{\pm}}_{\mathrm{AB}}
	&=
	\dfrac{1}{\sqrt{2}}
	\ket{01}_{\mathrm{AB}}
	\pm
	\dfrac{1}{\sqrt{2}}
	\ket{10}_{\mathrm{AB}}.\label{eq:Psi}
\end{align}
The duplex encoding transforms the initial Bell state $\ket{\Phi^+}_{\mathrm{AB}}$ to $\ket{\psi_1}_\mathrm{AB}$, one of the four Bell states $\ket{\Phi^{\pm}}_{\mathrm{AB}}$ and $\ket{\Psi^{\pm}}_{\mathrm{AB}}$.
\begin{figure*}[t!]
    \centering
    \subfigure[Type I]
    {
        \includegraphics[width=0.42\textwidth]{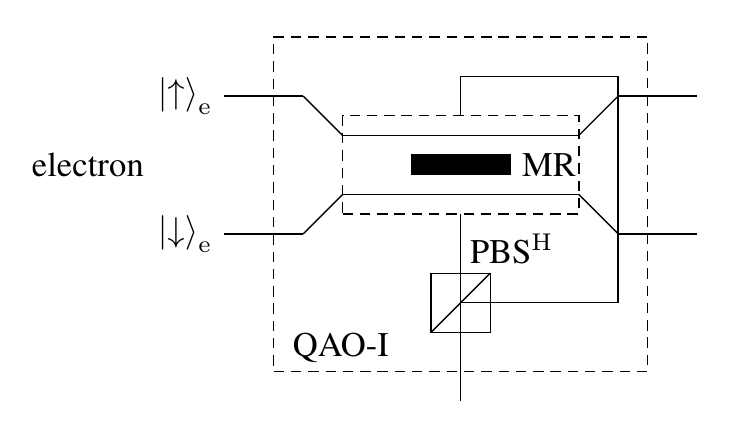}
        \label{fig:QAO:1}
    }
    \subfigure[Type II]
    {
        \includegraphics[width=0.42\textwidth]{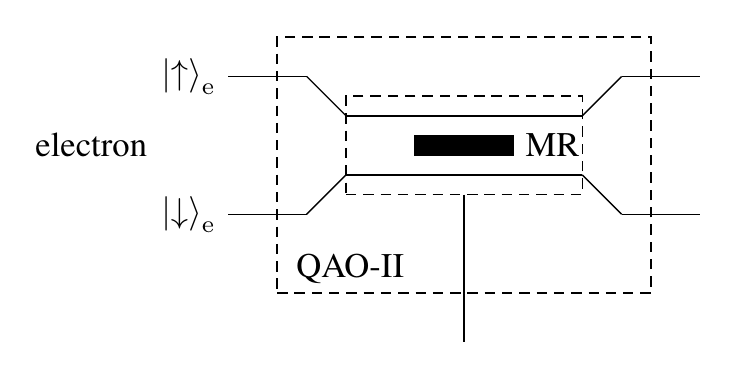}
        \label{fig:QAO:2}
    }

\caption{A quantum absorptive object (electron) for (a) the QZ gate (type I) and (b) the CQZ gate (type II). The electron takes the superposition of two paths $\ket{\uparrow}_{\mathrm{e}}$ and $\ket{\downarrow}_{\mathrm{e}}$. In type I, the electron states
$\ket{\uparrow}_\mathrm{e}$ and $\ket{\downarrow}_\mathrm{e}$ act as the presence (absence) state $\ket{1 \left(0\right)}_\mathrm{AO}$ and the absence (presence) state $\ket{0 \left(1\right)}_\mathrm{AO}$ of the absorptive object for the H(V)-QZ gate, respectively. In type II, the electron states simply act as $\ket{\uparrow}_\mathrm{e}=\ket{0}_\mathrm{AO}$ and $\ket{\downarrow}_\mathrm{e}=\ket{1}_\mathrm{AO}$ for the CQZ gate. If the photon is absorbed by the electron, the electron state is in an erasure state orthogonal to $\ket{\uparrow}_{\mathrm{e}}$ and $\ket{\downarrow}_{\mathrm{e}}$.
}
\label{fig:QAO}
\end{figure*}

To transfer the classical information in both directions simultaneously, Alice and Bob perform the \ac{DNOT} operation in a counterfactual way where Alice's qubit acts as a target bit and Bob's qubit acts as a control bit. The \ac{DNOT} operation \emph{disentangles} the encoded Bell state $\ket{\psi_1}_\mathrm{AB}$ to produce $\ket{\psi_2}_\mathrm{AB}$ as follows:
\begin{align}
\ket{\psi_2}_\mathrm{AB}:
\begin{cases}
\ket{\Phi^{\pm}}_{\mathrm{AB}}
	&
	\rightarrow
	\ket{0}_{\mathrm{A}}
	\ket{\pm}_{\mathrm{B}},\\
\ket{\Psi^{\pm}}_{\mathrm{AB}}
	&
	\rightarrow
	\ket{1}_{\mathrm{A}}
	\ket{\pm}_{\mathrm{B}},
\end{cases}
\end{align}
where $\ket{\pm}=\left(\ket{0}\pm\ket{1}\right)/\sqrt{2}$ is the Hadamard basis. 
To decode the classical information, Alice directly measures her qubit and decides the one-bit message $\rv{b}_2$, whereas Bob first applies the Hadamard gate $\M{H}$ followed by measuring his qubit in computational basis and decodes the one-bit message $\rv{b}_1$. Alice and Bob decide the decoded messages as $\rv{b}_2$ and $\rv{b}_1$ from their post-measurement states $\ket{\rv{b}_2}_{\mathrm{A}}$ and $\ket{\rv{b}_1}_{\mathrm{B}}$, respectively. Here it is important to note that whenever a physical particle is found in the quantum channel during the implementation of \ac{DNOT} operation, the protocol discards it and declares an erasure of the classical information $\rv{b}_1 \rv{b}_2$.
\begin{figure}[t!]
    \centering
        \includegraphics[width=0.55\textwidth]{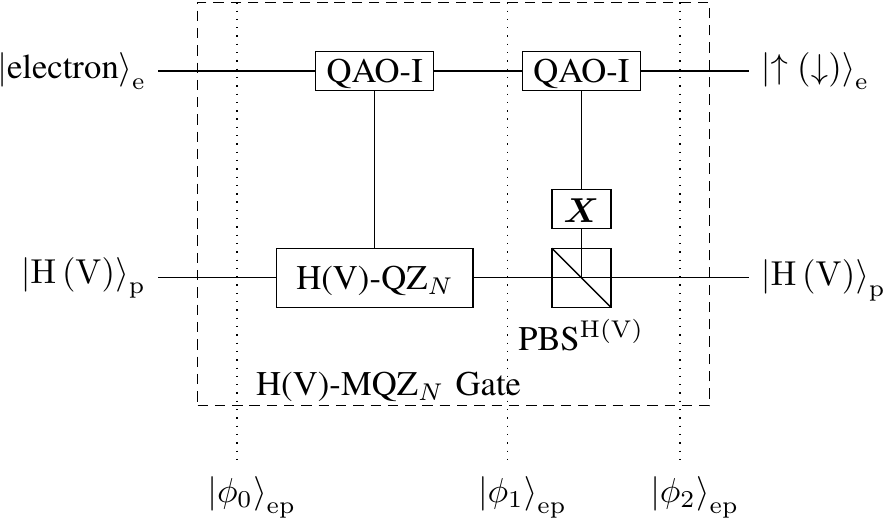}
    \caption{A H(V)-\ac{MQZ}$_N$ interaction where the superposition state $\ket{\text{electron}}_\mathrm{e}=\alpha \ket{\uparrow}_{\mathrm{e}}+\beta\ket{\downarrow}_{\mathrm{e}}$ of the quantum absorptive object (electron) is collapsed to $\ket{\uparrow\left(\downarrow\right)}_\mathrm{e}$ (dequantumization) using the H(V)-\ac{QZ}$_N$ gate unless the photon is absorbed by the electron. 
If the photon is found in the quantum channel, the pair of photon and electron is discarded in transforming $\ket{\phi_1}_\mathrm{ep}$ to $\ket{\phi_2}_\mathrm{ep}$ where the photon that has travelled over the channel is diverted again to the quantum absorptive object and absorbed by the electron. This electron-photon interaction is designed to output the photon and electron by using the presence state (blocking event) only. Hence, the protocol is fully counterfactual. }
        \label{fig:EPI-CQC}
\end{figure}

\subsection{\ac{MQZ} Duplex Coding}
In this section, the quantum duplex coding protocol is demonstrated using the H(V)-\ac{QZ}$_N$ gate. As shown in Fig.~\ref{fig:QAO}, an electron as a quantum absorptive object for duplexing coding takes the superposition of two paths $\ket{\uparrow}_{\mathrm{e}}$ and $\ket{\downarrow}_{\mathrm{e}}$ where the subscript $\mathrm{e}$ denotes the electron. In type I (Fig.~\ref{fig:QAO:1}), the electron state $\ket{\uparrow\left(\downarrow\right)}_\mathrm{e}$ or $\ket{\downarrow\left(\uparrow\right)}_\mathrm{e}$ acts as the presence state $\ket{1}_\mathrm{AO}$ or the absence state $\ket{0}_\mathrm{AO}$ of the absorptive object for the H(V)-QZ$_N$ gate. For counterfactuality, we setup the electron-photon interaction H(V)-\ac{MQZ}$_N$ shown in Fig.~\ref{fig:EPI-CQC} where the quantum absorptive object is in the superposition state

\begin{align}
\ket{\text{electron}}_\mathrm{e}
=
	\alpha
	\ket{\uparrow}_{\mathrm{e}}
	+
	\beta
	\ket{\downarrow}_{\mathrm{e}}\label{eq:electron}
\end{align}
with $|\alpha|^2+|\beta|^2=1$.
The H(V)-\ac{MQZ}$_N$ interaction collapses this quantum state by entangling and disentangling the electron-photon pair
\begin{align}
\ket{\phi_0}_\mathrm{ep}
=
	\ket{\text{electron}}_\mathrm{e}
	\ket{\mathrm{H}\left(\mathrm{V}\right)}_{\mathrm{p}}
\end{align} 
as follows:
\begin{align}	\label{eq:QHV:1}
\ket{\phi_0}_\mathrm{ep}
&\rightarrow
	\ket{\phi_1}_\mathrm{ep}
	=
	\alpha
	\ket{\uparrow\mathrm{H}}_{\mathrm{ep}}
	+
	\beta
	\ket{\downarrow\mathrm{V}}_{\mathrm{ep}} \\
&\rightarrow
	\ket{\phi_2}_\mathrm{ep}
	=
	\ket{\uparrow\left(\downarrow\right)}_\mathrm{e}
	\ket{\mathrm{H}\left(\mathrm{V}\right)}_{\mathrm{p}}	
\end{align}
unless the photon is absorbed by the electron, with probability 
\begin{align}
\left(
	1-
	\Delta_0
	\sin^2\theta_N
\right)^N
\Delta_0
\end{align}
where $\Delta_0=|\alpha|^2 \left(|\beta|^2\right)$ is the probability that the electron is in the presence state for the H(V)-QZ$_N$ gate.
\begin{figure*}[t!]
 \centering{
\includegraphics[width=0.9\textwidth]{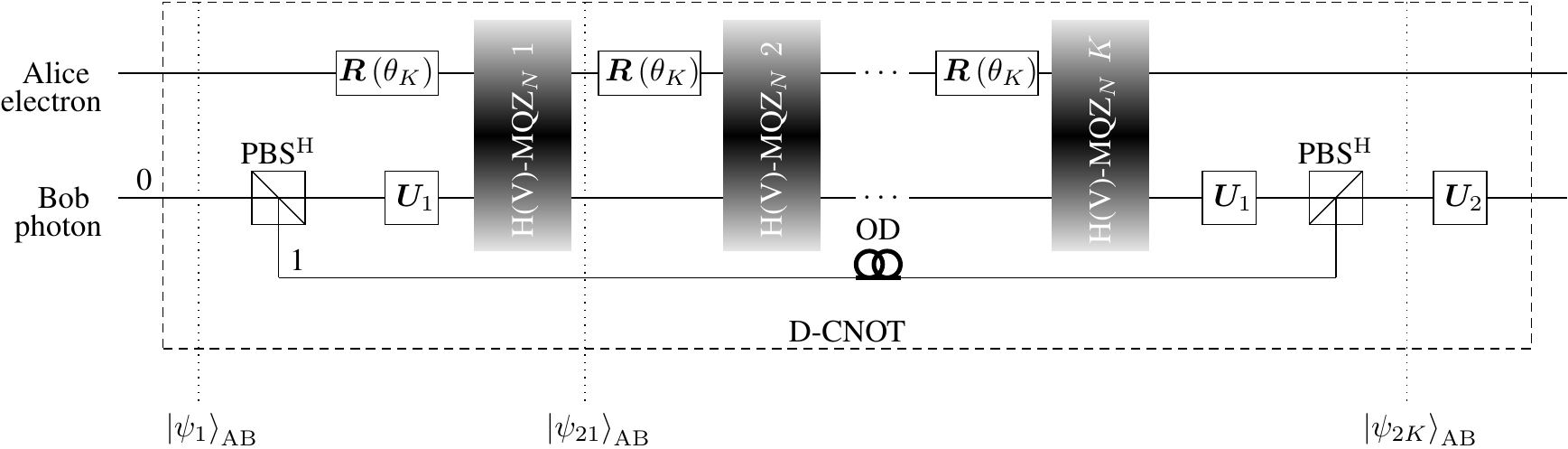}}
    \caption{
A \ac{MQZ} \ac{DNOT} operation for Bell states when $\rv{b}_2=0\left(1\right)$. Here, $\M{R}\left(\theta_K\right)$ is a rotation operator of rotation angle $\theta_{K}$ where $K$ is the number of H(V)-\ac{MQZ}$_N$ gates. Initially, Alice and Bob have the maximally entangled state $\ket{\psi_1}_\mathrm{AB}$, which is transformed by $K$ sets of the $\theta_{K}$ rotation and \ac{MQZ} gates successively to the separable state $\ket{\psi_{2K}}_\mathrm{AB}$ in a controlled manner. Finally, the $\M{U}_2$ operator is performed on the recombined photon to complete the \ac{MQZ} \ac{DNOT} operation. 
    }
    \label{fig:QZ-CQC}
\end{figure*}

The second (first) term of $\ket{\phi_1}_\mathrm{ep}$ is the outcome corresponding to the electron in the absence state for the H(V)-QZ$_N$ gate. Since this outcome is not counterfactual, it is discarded (absorbed) by the electron using the polarizing beam splitter PSB$^{\mathrm{H}\left(\mathrm{V}\right)}$ and 
the $\M{X}$ operator. Hence, whenever the photon is found in the quantum channel, the electron absorbs it and becomes in an erasure state, leading the  \ac{MQZ} gate to output no photon and electron (e.g., particles in the erasure state). To discard the factual (non-counterfactual) outcome $\ket{\mathrm{V}\left(\mathrm{H}\right)}_{\mathrm{p}}$ of the H(V)-QZ$_N$ gate, we can simply use a photon detector after PBS$^{\mathrm{H}\left(\mathrm{V}\right)}$. Instead, we redirect this photon component to the quantum absorptive object (followed by the $\sigma_x$ operator) 
to be absorbed by the electron. This enables the protocol to abort nonlocally by discarding both the photon and the electron whenever its counterfactuality is broken.

To implement the \ac{DNOT} operation for \ac{MQZ} duplex coding, we cascade $K$ H(V)-\ac{MQZ}$_N$ gates, where Alice is equipped with the electron and Bob has the \ac{MQZ} gates (see Fig.~\ref{fig:QZ-CQC}). Consider that
\begin{align}
\ket{0}_\mathrm{A}
	&=
	\ket{0}_\mathrm{e},\label{eq:A:0}\\
\ket{1}_\mathrm{A}
	&=
	\ket{1}_\mathrm{e},\\
\ket{0}_\mathrm{B}
	&=
	\ket{\mathrm{H}}_\mathrm{p},\\
\ket{1}_\mathrm{B}
	&=
	\ket{\mathrm{V}}_\mathrm{p}.\label{eq:B:1}
\end{align}
The \ac{MQZ} duplex coding protocol takes the following steps to implement \ac{DNOT} operation after encoding the classical information $\rv{b}_1\rv{b}_2$.
\begin{figure}[t!]
 \centering{
\includegraphics[width=0.65\textwidth]{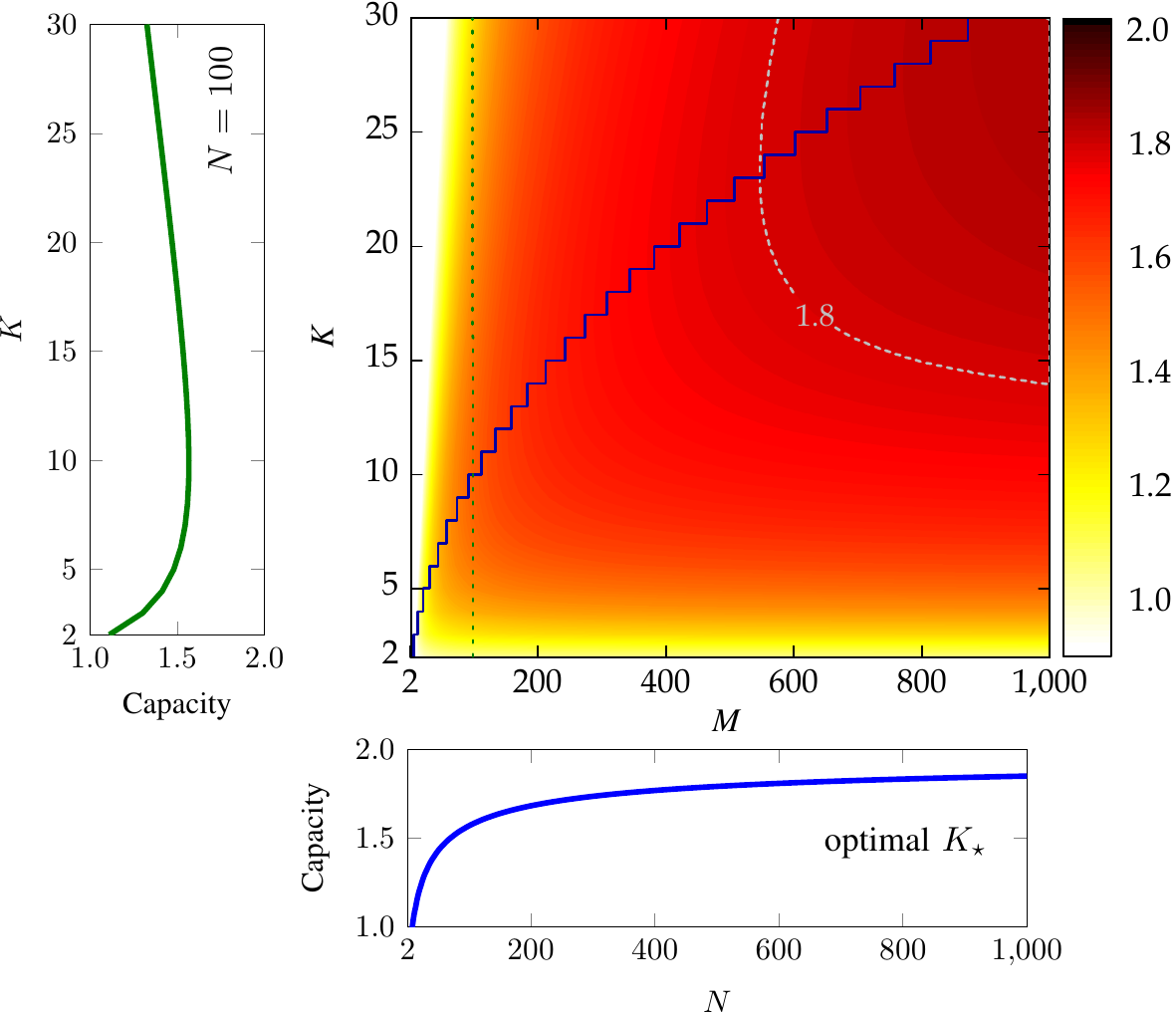}}
    \caption{
Capacity $C$ [bits/Bell-pair] of \ac{MQZ} duplex coding as a function of $N$ and $K$. Since the success probability $\zeta_\mathrm{c}$ in \eqref{eq:zeta-c} is concave in $K > 1$ for any positive integer $N$ (see the left plot), there exists the optimal value (positive integer) of $K$ that maximizes the capacity for a given $N$. 
The blue solid line is the trajectory of capacity  as a function of $N$ for the optimal values of $K$. The left plot depicts the capacity as a function of $K$ when $N=100$.  We also plot the trajectory of $\left(N,K\right)$ achieving the capacity of 1.8 bits/Bell-pair (white dashed line). 
    }
    \label{fig:CAP}
\end{figure}
\begin{enumerate}
\item 
Bob starts the protocol by separating the H and V components of the photon into two paths $\ket{0}_\mathrm{C}$ and $\ket{1}_\mathrm{C}$, respectively by using PBS$^\mathrm{H}$. Bob locally applies the unitary operation $\M{U}_1=\M{X}^{\rv{b}_2}$ on the component of the photon in path state $\ket{0}_\mathrm{C}$. 
%


\item 
Alice applies the rotation operation $\M{R}\left(\theta_K\right)$ on her qubit:  
\begin{align}
\M{R}\left(\theta_K\right)
	=
	\renewcommand*{\arraystretch}{1.1}
	\begin{bmatrix}
	\cos\theta_K
	&
	-
	\sin\theta_K\\
	\sin\theta_K
	&
	\cos\theta_K
	\end{bmatrix},
\end{align}
where $\iota=\sqrt{-1}$ and $\theta_K=\pi/\left(2K\right)$. The rotation gate $\M{R}\left(\theta_K\right)$ transforms $\ket{0}_{\mathrm{A}}$ and $\ket{1}_{\mathrm{A}}$ as follows:
\begin{align}
\ket{0}_{\mathrm{A}}
&\rightarrow
	\cos\theta_K
	\ket{0}_{\mathrm{A}}
	+
	\sin\theta_K
	\ket{1}_{\mathrm{A}}\\
\ket{1}_{\mathrm{A}}
&\rightarrow
	\cos\theta_K
	\ket{1}_{\mathrm{A}}
	-\sin\theta_K
	\ket{0}_{\mathrm{A}}.
\end{align}

\item 
Bob inputs the component of the photon in path $\ket{0}_\mathrm{C}$ of the photon to H(V)-\ac{MQZ}$_N$ gate for $\rv{b}_2=0\left(1\right)$. Unless the photon is absorbed by the electron, it transforms the composite state of the Alice and Bob to $\ket{\psi_{21}}_\mathrm{AB}$ 
\begin{align}
\ket{\psi_{21}}_\mathrm{AB}:&
\begin{cases}
\ket{\Phi^\pm}_\mathrm{ABC}
\rightarrow
	\frac{1}{\sqrt{2}}
	\bigl(
		\ket{000}_\mathrm{ABC}
		\pm
		\cos\theta_K
		\ket{111}_\mathrm{ABC}
		\mp
		\sin\theta_K
		\ket{011}_\mathrm{ABC}
	\bigr), 
\\
\ket{\Psi^\pm}_\mathrm{ABC}
\rightarrow
	\frac{1}{\sqrt{2}}
	\bigl(
		\pm
		\ket{110}_\mathrm{ABC}
		+
		\cos\theta_K
		\ket{011}_\mathrm{ABC}
		+
		\sin\theta_K
		\ket{111}_\mathrm{ABC}
	\bigr),
\end{cases}\label{eq:first CEPI gate}
\end{align}
with probability
\begin{align}
\begin{aligned}
\lambda_2
&=
\Big(
	1-
	\frac{1}{2}
	\cos^2\theta_K
	\sin^2\theta_{N}
\Big)^{N} 
\Big(
	1-
	\frac{1}{2}
	\sin^2\theta_K
\Big),
\end{aligned}
\end{align}
which tends to one as $N,K\rightarrow\infty$. Whenever the physical particle is traveled over the quantum channel between Alice and Bob, it is absorbed by the electron and the protocol declares an erasure. 

\item 
Alice and Bob keep recurring the step 2) and step 3) for remaining $K-1$ H(V)-\ac{MQZ}$_N$ gates unless the protocol declares the erasure with probability $1-\zeta_c$ where 
\begin{align} \label{eq:zeta-c}
\zeta_\mathrm{c}
=
	\lambda_2^K.
\end{align}

\begin{table}[t!]
\centering
\setlength{\tabcolsep}{8pt}
{\renewcommand{\arraystretch}{1.5}
\caption{Decoding the classical message $\rv{b}_1\rv{b}_2$ for the \ac{MQZ} dyplex coding.}\label{tab:DCM}
\vspace{0.2cm}
\begin{tabular}{P{1.4cm} | P{1.2cm} | P{1.2cm} | P{1.2cm}}
\hline
\multicolumn{2}{c|}{Alice} & \multicolumn{2}{c}{Bob}\\
\hline
Electron & $\rv{b}_2$ & Photon & $\rv{b}_1$ \\ 
\hline
\hline 
 $\ket{\uparrow}_{\mathrm{e}}$ & 0 & $\ket{\mathrm{H}}_{\mathrm{p}}$ & 0\\
 $\ket{\downarrow}_{\mathrm{e}}$ & 1 & $\ket{\mathrm{V}}_{\mathrm{p}}$ & 1\\ 
\hline
\end{tabular}}
\end{table}

\item
Bob applies $\M{U}_1$ and recombines the H and V components of the photon. The encoded Bell pair $\ket{\psi_1}_\mathrm{AB}$ is disentangled to $\ket{\psi_{2K}}_\mathrm{ABC}$ as follows:
\begin{align}
\ket{\psi_{2K}}_\mathrm{ABC}:
	\begin{cases}
	\ket{\Phi^\pm}_\mathrm{ABC}
	\rightarrow
 	\ket{0}_\mathrm{A}
	\ket{\mp}_\mathrm{B}
	\ket{0}_\mathrm{C}, \\
	\ket{\Psi^\pm}_\mathrm{ABC}
	\rightarrow
	\ket{1}_\mathrm{A}
	\ket{\pm}_\mathrm{B}
	\ket{0}_\mathrm{C}.
	\end{cases} 
\end{align}

\item 
Bob finally performs the $\M{U}_2=\M{Z}^{1-\rv{b}_2}$ operation on the component of the photon in path state $\ket{0}_\mathrm{C}$ to complete the \ac{MQZ} \ac{DNOT} operation   
\end{enumerate}

Alice measures the path of the electron to decode the classical information $\rv{b}_2$. Bob first applies the Hadamard gate $\M{H}$ to the photon, which transforms its polarization as 
\begin{align}
\M{H}
\ket{+}_\mathrm{B}
&\rightarrow
	\ket{\mathrm{H}}_\mathrm{p}, \\ 
\M{H}
\ket{-}_\mathrm{B}
&\rightarrow
	\ket{\mathrm{V}}_\mathrm{p}. 
\end{align}
Bob measures the  polarization of the existing photon to decode the classical information $\rv{b}_1$. Table~\ref{tab:DCM} shows the decoded messages corresponding to the measurement outcomes. The \ac{MQZ} duplex coding creates a full-duplex form of the classical \ac{BEC} with the erasure probability $1-\zeta_\mathrm{c}$ (see~\eqref{eq:zeta-c}). The bidirectional capacity $C$ in bits/Bell-pair of the \ac{MQZ} duplex coding is given by\footnote{This protocol enables each party to achieve unidirectional capacity of $\zeta_c$ bits/Bell-pair in each direction simultaneously by using only one Bell-pair.} 
\begin{align}
C
=
	2\zeta_\mathrm{c}
\end{align}
which tends to 2 bits/Bell-pair as $N,K \rightarrow \infty$ (see Fig.~\ref{fig:CAP}).

%
%
\begin{figure*}[t!]	
  \centering
\includegraphics[width=0.9\textwidth]{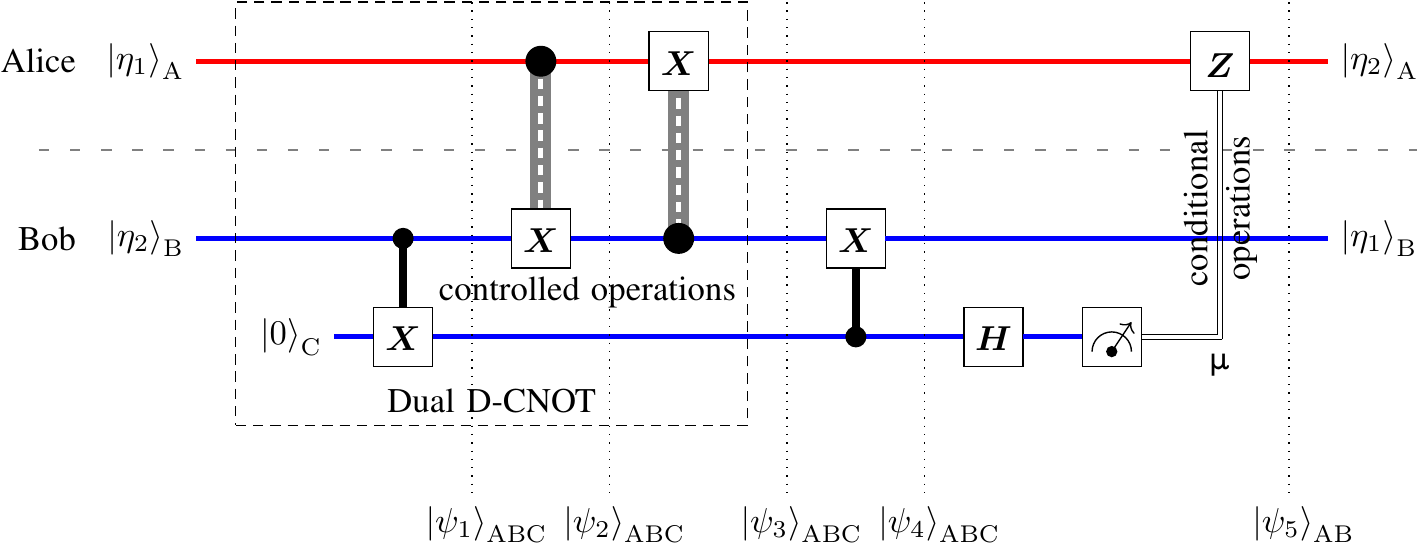}
    \caption{
Quantum telexchaning for quantum information $\ket{\eta_1\eta_2}_{\mathrm{AB}}$. Alice and Bob have an untangled pair of qubits $\ket{\eta_1}_{\mathrm{A}}=\alpha\ket{0}_{\mathrm{A}}+\beta\ket{1}_{\mathrm{A}}$ and $\ket{\eta_2}_{\mathrm{B}}=\gamma\ket{0}_{\mathrm{B}}+\delta\ket{1}_{\mathrm{B}}$ to communicate with each other. Bob starts the dual \ac{DNOT} protocol by entangling his message $\ket{\eta_2}_{\mathrm{B}}$ and ancilla $\ket{0}_{\mathrm{C}}$ with the local CNOT operation. Alice and Bob perform the nonlocal operations on their composite state, which entangles and disentangles these remote parties successively to exchange their quantum information counterfactually. Then, Bob and Alice perform local operations to decode each quantum message. Specifically, Bob performs the CNOT operation followed by the Hadamard gate $\M{H}$ to decode Alice's quantum message as $\ket{\eta_1}_{\mathrm{B}}=\alpha\ket{0}_{\mathrm{B}}+\beta\ket{1}_{\mathrm{B}}$. Bob then announces his ancilla measurement $\rv{\mu} \in \left\{0,1\right\}$ to Alice by classical communication. Using Bob's  announcement, Alice finally performs the $\M{Z}^{\rv{\mu}}$ operator on her qubit to decode Bob's quantum message as $\ket{\eta_2}_{\mathrm{A}}=\gamma\ket{0}_{\mathrm{A}}+\delta\ket{1}_{\mathrm{A}}$. 
      }
    \label{fig:QQDC}
\end{figure*}

\section{Quantum Telexchanging}\label{sec 4}

In this section, we develop a full-duplex quantum protocol  to transfer quantum information in both directions simultaneously and counterfactually. 

\subsection{Protocol}
Consider that Alice and Bob want to exchange their quantum states $\ket{\eta_1}_{\mathrm{A}}$ and $\ket{\eta_2}_{\mathrm{B}}$ simultaneously where
\begin{align}
\ket{\eta_1}_{\mathrm{A}}
	&=
	\alpha\ket{0}_{\mathrm{A}}
	+
	\beta\ket{1}_{\mathrm{A}},\label{eq:A}\\
\ket{\eta_2}_{\mathrm{B}}
	&=
	\gamma\ket{0}_{\mathrm{B}}
	+
	\delta\ket{1}_{\mathrm{B}}.\label{eq:B}
\end{align}

To transfer the quantum information in both directions at the same time, Alice and Bob perform the \ac{DDNOT} operation on their message qubits to entangle and disentangle them counterfactually.
Bob starts the protocol by entangling his qubit $\ket{\eta_2}_{\mathrm{B}}$ with his ancillary qubit $\ket{0}_{\mathrm{C}}$ by performing the CNOT operation locally as shown in Fig.~\ref{fig:QQDC}. Then, Alice and Bob have the separable composite state $\ket{\psi_1}_{\mathrm{ABC}}$ as follows:
\begin{align}  \label{eq:psi_1}
\ket{\psi_1}_{\mathrm{ABC}}
	=
	\ket{\eta_1}_\mathrm{A}
	\left(
		\gamma\ket{00}_{\mathrm{BC}}
		+
		\delta\ket{11}_{\mathrm{BC}}
	\right). 
\end{align}

Alice and Bob perform the two nonlocal CNOT operations on their qubits. In the first CNOT operation, Alice's message qubit acts as a control qubit and Bob's message qubit acts as a target qubit. In the second CNOT operation, Alice's message qubit acts as a target qubit and Bob's message qubit acts as a control qubit. These nonlocal operations transform the composite state $\ket{\psi_1}_{\mathrm{ABC}}$  
as in follows
\begin{align}
	 \label{eq:psi_2}
\ket{\psi_1}_{\mathrm{ABC}}
\rightarrow
	\ket{\psi_2}_{\mathrm{ABC}}
	&=
		\alpha\gamma
		\ket{000}_{\mathrm{ABC}} +
		\alpha\delta
		\ket{011}_{\mathrm{ABC}} +
		\beta\gamma
		\ket{110}_{\mathrm{ABC}} +
		\beta\delta
		\ket{101}_{\mathrm{ABC}}	\\
\rightarrow
	\ket{\psi_3}_{\mathrm{ABC}}
	&=
		\alpha\gamma
		\ket{000}_{\mathrm{ABC}} +
		\alpha\delta
		\ket{111}_{\mathrm{ABC}} +
		\beta\gamma
		\ket{010}_{\mathrm{ABC}} +
		\beta\delta
		\ket{101}_{\mathrm{ABC}}		\\
	&
	=
		\gamma\ket{00}_\mathrm{AC}
		\left(\alpha\ket{0}_\mathrm{B}+\beta\ket{1}_\mathrm{B}\right)+
		\delta\ket{11}_\mathrm{AC}
		\left(\alpha\ket{1}_\mathrm{B}+\beta\ket{0}_\mathrm{B}\right).\label{eq:psi_3}
\end{align}
Bob then applies the CNOT gate locally on his message and ancilla qubits to decode Alice's message state. It transforms $\ket{\psi_3}_{\mathrm{ABC}}$ as follows: 
\begin{align}
\begin{aligned}
\ket{\psi_4}_{\mathrm{ABC}}
	&=
		\left(
			\gamma\ket{00}_\mathrm{AC}+
			\delta\ket{11}_\mathrm{AC}
		\right)
		\left(
			\alpha\ket{0}_\mathrm{B}+
			\beta\ket{1}_\mathrm{B}
		\right).
 \end{aligned}
 \end{align}
To further disentangle Bob's ancilla and Alice's qubit, Bob applies the Hadamard gate $\M{H}$ on his ancilla followed by measuring it in computational basis. 
Bob announces this measurement outcome $\rv{\mu} \in \left\{0,1\right\}$ to Alice by classical communication and Alice finally performs the $\M{Z}^{\rv{\mu}}$ operation on her qubit to decode Bob's message state as follows: 
\begin{align}
\begin{aligned}
\ket{\psi_5}_{\mathrm{AB}}
	&=
		\left(
			\gamma\ket{0}_\mathrm{A}+
			\delta\ket{1}_\mathrm{A}
		\right)
		\left(
			\alpha\ket{0}_\mathrm{B}+
			\beta\ket{1}_\mathrm{B}
		\right)\\
	&=
		\ket{\eta_2}_\mathrm{A}
		\ket{\eta_1}_\mathrm{B}.
 \end{aligned}
 \end{align}
Whenever a physical particle is found in the quantum channel for the \ac{DDNOT} operation, the protocol discards it and declares an erasure of the quantum information $\ket{\eta_1\eta_2}_{\mathrm{AB}}$.
\begin{figure}[t!]
 \centering{\includegraphics[width=0.6\textwidth]{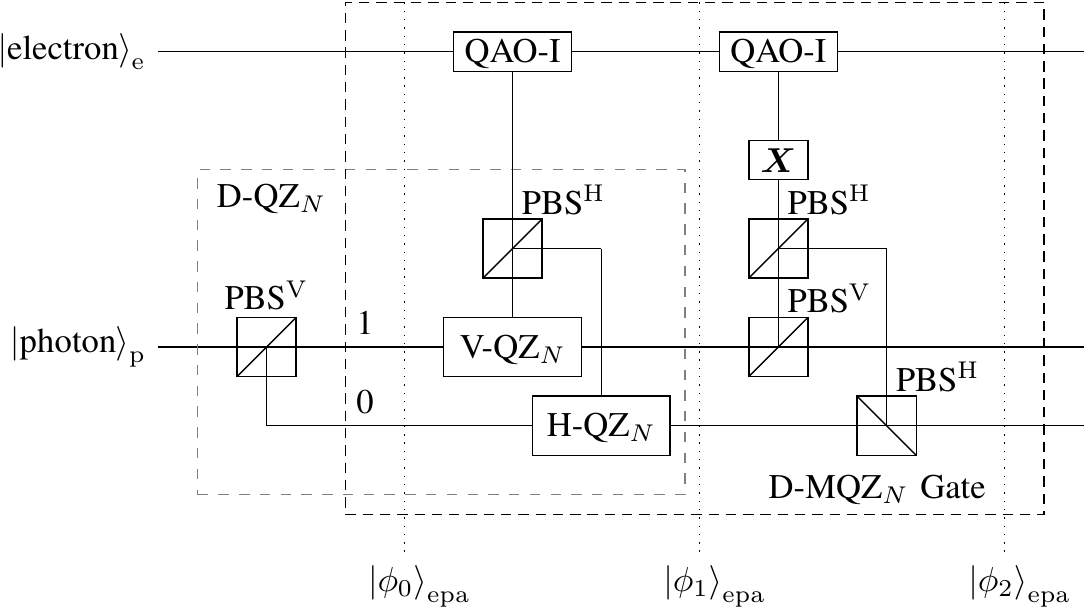}}
    \caption{
A \ac{DMQZ}$_N$ interaction where the quantum absorptive object (electron) gets entangled with the existing photon (unless absorbed by the electron) using the \ac{DQZ}$_N$ gate. Initially, the photon is in the superposition state $\ket{\text{photon}}_\mathrm{p}=\gamma\ket{0}_{\mathrm{p}}+\delta\ket{1}_{\mathrm{p}}$, which is entangled with the ancillary path state by PBS$^\mathrm{H}$ as $\ket{\text{photon}}_\mathrm{pa} = \gamma\ket{\mathrm{H0}}_{\mathrm{pa}} + \delta\ket{\mathrm{V1}}_{\mathrm{pa}}$ to start the \ac{DMQZ}$_N$ interaction. Similar to the \ac{MQZ} gate in Fig.~\ref{fig:EPI-CQC}, the \ac{DMQZ}$_N$ gate then transforms the electron-photon pair $\ket{\phi_0}_{\mathrm{epa}}$ to $\ket{\phi_2}_{\mathrm{epa}}=\gamma\ket{\uparrow\mathrm{H}0}_{\mathrm{epa}}+\delta\ket{\downarrow\mathrm{V}1}_{\mathrm{epa}}$ by using the blocking event only (unless the photon is absorbed by the electron).    
}
    \label{fig:EPI-QC}
\end{figure}

\subsection{\ac{MQZ}-\ac{CQZ} Telexchanging}
In this section, the quantum telexchanging protocol is demonstrated using the H(V)-\ac{QZ}$_N$ and H(V)-\ac{CQZ}$_{M,N}$ gates. Fig.~\ref{fig:EPI-QC} shows the dual form of the \ac{MQZ} gate in Fig.~\ref{fig:EPI-CQC}. For counterfacuality, this \ac{DMQZ}$_N$ gate works similarly to the H(V)-\ac{MQZ}$_N$ gate. The only difference is that the superposition polarization state $\ket{\text{photon}}_\mathrm{p}=\gamma\ket{\mathrm{H}}_{\mathrm{p}}+\delta\ket{\mathrm{V}}_{\mathrm{p}}$ of the input photon is entangled with the ancillary path state in the \ac{DMQZ}$_N$ gate as follows:  
\begin{align}
\ket{\text{photon}}_\mathrm{pa}
=
	\gamma\ket{\mathrm{H0}}_{\mathrm{pa}}
	+
	\delta\ket{\mathrm{V1}}_{\mathrm{pa}}	
\end{align}
where the ancilla states $\ket{0}_\mathrm{a}$ and $\ket{1}_\mathrm{a}$ show the paths for the H- and V-QZ gates, respectively.  The \ac{DMQZ}$_N$ gate transforms the electron-photon pair
\begin{align}
\ket{\phi_0}_\mathrm{epa}
=
	\ket{\text{electron}}_\mathrm{e}
	\ket{\text{photon}}_\mathrm{pa}
\end{align} 
as follows:
%
\begin{align}	
\ket{\phi_0}_\mathrm{epa}
&\rightarrow
	\ket{\phi_1}_\mathrm{epa}
	=
	\alpha
	\gamma
	\ket{\uparrow\mathrm{H}0}_{\mathrm{epa}}
	+
	\beta
	\gamma
	\ket{\downarrow\mathrm{V}0}_{\mathrm{epa}}
	+
	\alpha
	\delta
	\ket{\uparrow\mathrm{H}1}_{\mathrm{epa}}
	+
	\beta
	\delta
	\ket{\downarrow\mathrm{V}1}_{\mathrm{epa}}
\\
&\rightarrow
	\ket{\phi_2}_\mathrm{epa}
	=
	\gamma
	\ket{\uparrow\mathrm{H}0}_{\mathrm{epa}}	
	+
	\delta
	\ket{\downarrow\mathrm{V}1}_{\mathrm{epa}}	
\end{align}
unless the photon is absorbed by the electron, with probability 
\begin{align}
\left(
	1-
	\Delta_1
	\sin^2\theta_N
\right)^N
\Delta_1
\end{align}
where $\Delta_1=\vert\alpha\gamma\vert^2+\vert\beta\delta\vert^2$ is the probability that the electron is in the presence state for the QZ gates in both paths. 

\begin{figure*}[t!]	
  \centering
\includegraphics[width=0.935\textwidth]{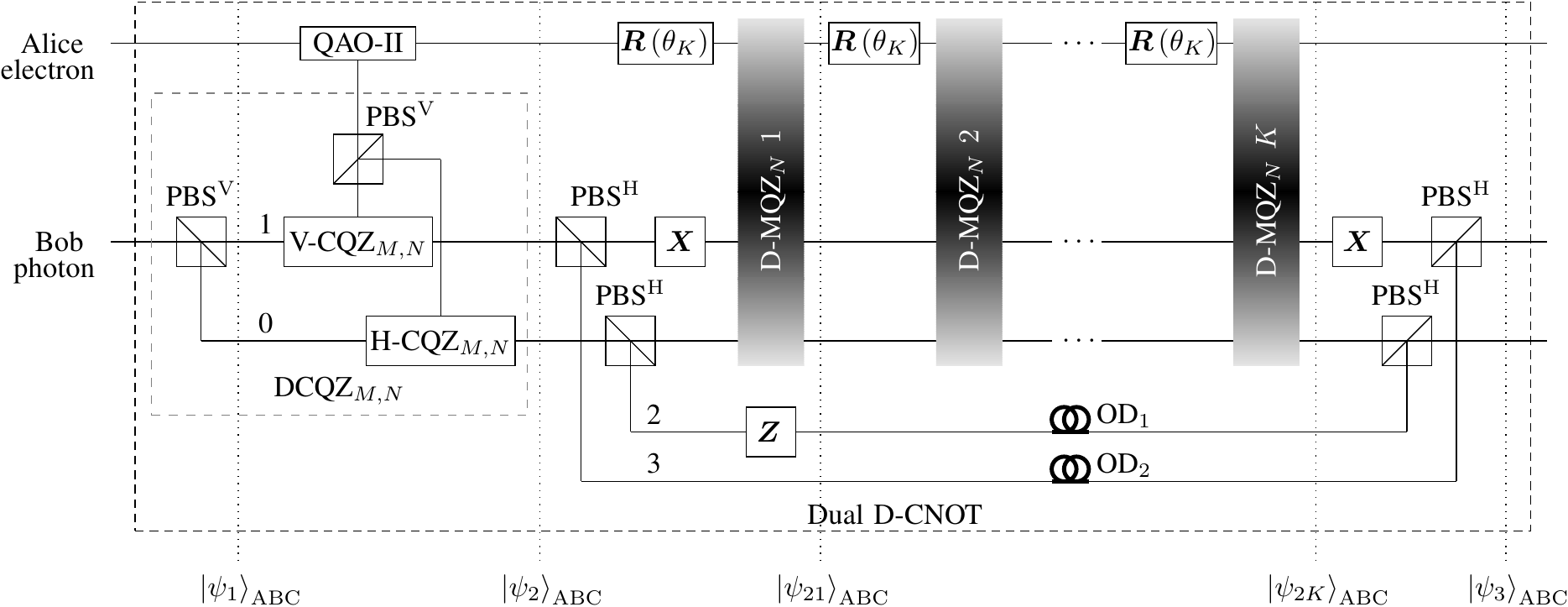}
    \caption{
A \ac{MQZ}-\ac{CQZ} \acf{DDNOT} operation for an unknown pair of quantum states. Initially, Alice and Bob have an untangled pair $\ket{\eta_1\eta_2}_{\mathrm{AB}}$ of the electron and photon. This message pair is entangled by the DCQZ$_{M,N}$ gate and disentangled by $K$ rounds of the $\theta_K$ rotation and \ac{DMQZ}$_N$ gates counterfactually in a controlled manner. Finally, Bob applies the $\M{X}$ on the photon component in path $\ket{1}_{\mathrm{C}}$ and recombines the respective components of the photon to complete the \ac{MQZ}-\ac{CQZ} \ac{DDNOT} operation.  
      }
    \label{fig:QZ-QDC}
\end{figure*}

For \ac{MQZ}-\ac{CQZ} quantum telexchaning, Alice and Bob initially have an untangled pair of quantum information $\ket{\eta_1}_{\mathrm{A}}$ and $\ket{\eta_2}_{\mathrm{B}}$ prepared in the electron and photon, respectively,  
where we consider  \eqref{eq:A:0}--\eqref{eq:B:1} again. Bob starts the protocol by throwing his photon towards PBS$^{\mathrm{H}}$ to entangle the polarization state $\ket{\text{photon}}_{\mathrm{B}}$ with the ancillary path state $\ket{0}_\mathrm{C}$ as shown in Fig.~\ref{fig:QZ-QDC}. Then, Alice and Bob have the composite state $\ket{\psi_1}_{\mathrm{ABC}}$ in \eqref{eq:psi_1}. To implement the \ac{DDNOT} operation in Fig.~\ref{fig:QQDC}, Alice and Bob first  entangle their qubits counterfactually by using the DCQZ$_{M,N}$ gate and then perform $K$ \ac{DMQZ}$_N$ gates for controlled disentanglement (see Fig.~\ref{fig:QZ-QDC}).  The \ac{MQZ}-\ac{CQZ} telexchanging takes the following steps to device the  \ac{DDNOT} operation after preparing the message states. 
\begin{figure}[t!]
 \centering{
\includegraphics[width=0.5\textwidth]{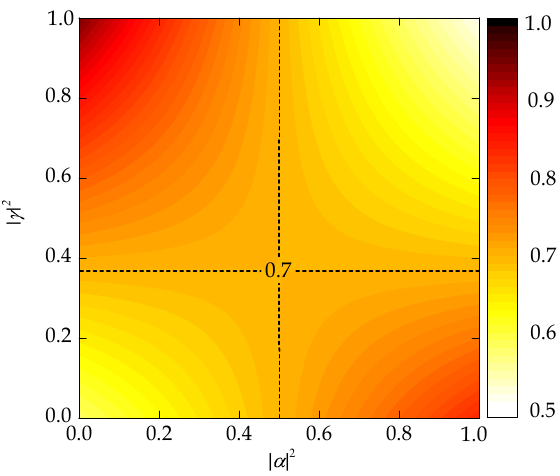}}
    \caption{
Transfer efficiency (fidelity) $\zeta_\mathrm{q}$ for the \ac{MQZ}-\ac{CQZ} telexchaning as a function of $\vert\alpha\vert^2$ and $\vert\gamma\vert^2$ when $N=100$ and $M_\star=K_\star=10$ where $M_\star$ and $K_\star$ are the optimal values that maximize $\zeta_\mathrm{q}$ for given $N$ such that $M_\star = \arg \max_M \zeta_3$ and $K_\star = \arg \max_K \zeta_4^K$. When $\vert\alpha\vert^2=1/2$, the transfer efficiency is equal to $\zeta_\mathrm{q}=0.659$ independent of the message states (black dashed line). We can see that $\zeta_\mathrm{q}$ increases as $\Delta_1 \rightarrow 0$ (the message states are collapsing to the classical information). When $N=100$, the maximum efficiency is equal to $\zeta_\mathrm{q}=0.903$ for $\vert\alpha\vert^2=0$ and $\vert\gamma\vert^2=1$.
    }
    \label{fig:Fidelity}
\end{figure}
\begin{enumerate}
\item
Alice and Bob start the dual \ac{DNOT} protocol by entangling their message states $\ket{\text{electron}}_{\mathrm{A}}$ and $\ket{\text{photon}}_{\mathrm{B}}$ counterfactually where Alice's message state $\ket{\text{electron}}_{\mathrm{A}}$ acts as a quantum \ac{AO} and Bob is equipped with the \ac{DCQZ} gate. Unless the photon is absorbed by the electron or discarded at the detector in the \ac{DCQZ}$_{M,N}$ gate, this counterfactual entanglement transforms the encoded state $\ket{\psi_1}_{\mathrm{ABC}}$ to $\ket{\psi_2}_{\mathrm{ABC}}$ in \eqref{eq:psi_2}, with probability 
\begin{align}	
\begin{aligned}
\label{eq:lambda:3}
\lambda_3
&=
	\left(1-\vert\alpha\vert^2\sin^2\theta_M\right)^M
	\prod_{m=1}^M
	\left[
		1-
		\vert\beta\vert^2\sin^2\left(m\theta_M\right)
		\sin^2\theta_N
	\right]^{N},
\end{aligned}
\end{align}
which tends to one as $M,N \rightarrow \infty$ \cite{ZJS:18:SR}.

\item 
Bob applies PBS$^\mathrm{H}$ in each path of the photon to separate the H and V components of the photon  and performs the $\M{U}_3$ operation locally where 
\begin{align}
\begin{aligned}
\M{U}_3
	&=
	\M{I}
	\otimes
	\left(
		\ket{0}_{\text{C}}
		\hspace{-0.065cm}
		\bra{0}
		+
		\ket{3}_{\text{C}}
		\hspace{-0.065cm}
		\bra{3}
	\right)
	+
	\M{X}
	\otimes
	\ket{1}_{\text{C}}
	\hspace{-0.065cm}
	\bra{1}
	+
	\M{Z}
	\otimes
	\ket{2}_{\text{C}}
	\hspace{-0.065cm}
	\bra{2}.
\end{aligned}
\end{align}

\item 
Alice applies the rotation operation $\M{R}\left(\theta_K\right)$ on her qubit and Bob applies the \ac{DMQZ} gate on the components of the photon in path state $\ket{0}_\mathrm{C}$ and $\ket{1}_\mathrm{C}$. Unless the photon is discarded in the \ac{DMQZ} gate, the \ac{DMQZ} gate transforms $\ket{\psi_2}_\mathrm{ABC}$ as follows  
\begin{align}
\begin{aligned}
\ket{\psi_{21}}_{\mathrm{ABC}}
&=
	\alpha
	\gamma
	\ket{000}_{\mathrm{ABC}}
	+
	\beta
	\delta
	\ket{111}_{\mathrm{ABC}}
	+
	\beta
	\gamma
	\sin\theta_K
	\ket{012}_{\mathrm{ABC}}
	\\
&\quad-
	\beta
	\gamma
	\cos\theta_K
	\ket{112}_{\mathrm{ABC}}	
	+
	\alpha
	\delta
	\cos\theta_K
	\ket{013}_{\mathrm{ABC}}	
	+
	\alpha
	\delta
	\sin\theta_K
	\ket{113}_{\mathrm{ABC}},	
\end{aligned} 
\end{align}  
with probability
\begin{align}
\begin{aligned} 	\label{eq:lambda:4}
\lambda_4
	&=
	\Bigl(
		1
		-
		\Delta_1
		\cos^2\theta_K
		\sin^2\theta_N
	\Bigr)^N
	\Bigl(
		1
		-
		\Delta_1
		\sin^2\theta_K
	\Bigr),
\end{aligned} 
\end{align}  
which tends to one as $N,K\rightarrow\infty$.

\item
Alice and Bob keep repeating the step 3) for remaining  $K-1$ \ac{DMQZ} gates. The remaining $K-1$ \ac{DMQZ} gates transform the composite state $\ket{\psi_{21}}$ to $\ket{\psi_{2K}}$ as follows: 
\begin{align}
\begin{aligned}
\ket{\psi_{2K}}_{\mathrm{ABC}}
&=
	\alpha
	\gamma
	\ket{000}_{\mathrm{ABC}}
	+
	\beta
	\delta
	\ket{111}_{\mathrm{ABC}}
	+
	\beta
	\gamma
	\ket{012}_{\mathrm{ABC}}
	+
	\alpha
	\delta
	\ket{113}_{\mathrm{ABC}},
\end{aligned}
\end{align} 
unless the photon is discarded in the \ac{DDNOT} operation with probability $1-\zeta_\mathrm{q}$ where
\begin{align}  
\zeta_\mathrm{q}
=
	\lambda_3
	\lambda_4^K.
\end{align}
Whenever the physical particle is transmitted over the quantum channel between Alice and Bob, the protocol declares an erasure of the quantum message $\ket{\eta_1\eta_2}_\mathrm{AB}$. 

\item 
Bob applies $\M{U}_4$ on his photon and recombines the  respective components of the photon to complete the \ac{DDNOT} operation $\left(\ket{\psi_3}_\mathrm{ABC}\right)$ where
\begin{align}
\begin{aligned}
\M{U}_4
&=
	\M{I}
	\otimes
	\left(
		\ket{0}_{\text{C}}
		\hspace{-0.065cm}
		\bra{0}
		+
		\ket{2}_{\text{C}}
		\hspace{-0.065cm}
		\bra{2}
		+
		\ket{3}_{\text{C}}
		\hspace{-0.065cm}
		\bra{3}
	\right)
	+
	\M{X}
	\otimes
	\ket{1}_{\text{C}}
	\hspace{-0.065cm}
	\bra{1}.
\end{aligned}
\end{align}

\end{enumerate}
This \ac{MQZ}-\ac{CQZ} telexchaning for quantum information creates a full-duplex form of the \ac{QEC} \cite{BDS:97:PRL} with the erasure probability $1-\zeta_\mathrm{q}$ as follows: 
\begin{align}
\begin{aligned}
\ket{\eta_1\eta_2}_{\mathrm{AB}}
&\rightarrow
	\mathcal{N}
		\left(
			\ket{\eta_1\eta_2}_{\mathrm{AB}}
		\right)
		=
	\zeta_\mathrm{q}
	\ket{\eta_1\eta_2}_{\mathrm{BA}}
	\hspace{-0.12cm}			
	\bra{\eta_1\eta_2}+
	\left(
		1-\zeta_\mathrm{q}
	\right)
	\ket{\varepsilon}_\mathrm{BA}
	\hspace{-0.12cm}
	\bra{\varepsilon}
\end{aligned}
\end{align}
where $\mathcal{N}$ denotes the full-duplex \ac{QEC} formed by the protocol and $\ket{\varepsilon}_{\mathrm{BA}}$ is the erasure state orthogonal and independent to the message state $\ket{\eta_1\eta_2}_{\mathrm{AB}}$.
The transfer efficiency $\zeta_\mathrm{q}$ can be also viewed as the fidelity 
%
\begin{align}
F
=
	\bra{\eta_1\eta_2}
	\mathcal{N}
	\ket{\eta_1\eta_2}_{\mathrm{AB}},
\end{align}
which depends on the message state  $\ket{\eta_1\eta_2}_{\mathrm{AB}}$ in general.
Since $\Delta_1 = 1/2$ if $|\alpha|^2=|\beta|^2=1/2$, this dependence vanishes when Alice's message $\ket{\eta_1}_{\mathrm{A}}$ is in the superposition state of equiprobable $\ket{0}_\mathrm{A}$ and $\ket{1}_\mathrm{A}$ (see Fig.~\ref{fig:Fidelity}).   
The quantum capacity $Q$ in qubits/electron-photon for the \ac{MQZ}-\ac{CQZ} telexchaning is given by
\begin{align}
Q=	
	2
	\max
	\left\{
		0,
		2\zeta_\mathrm{q}-1
	\right\}\label{eq: quantum capacity}
\end{align}
tending to 2 qubits/electron-photon as $K,M,N \rightarrow \infty$ (see Fig.~\ref{fig:QDC-QC-CAP}). 
\begin{figure}[t!]
 \centering{
\includegraphics[width=0.58\textwidth]{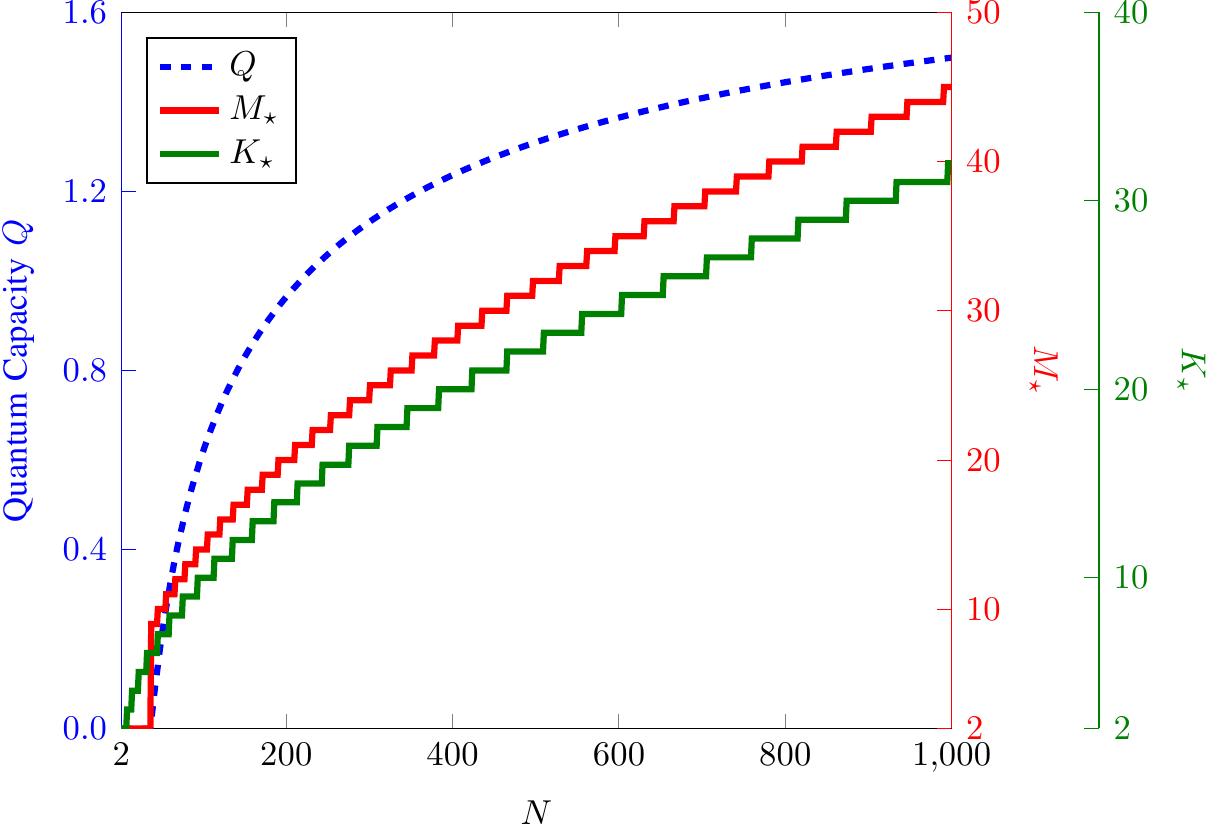}}
    \caption{
Quantum capacity $Q$ [qubits/electron-photon], $M_\star$, and $K_\star$ for the \ac{MQZ}-\ac{CQZ} telexchaning as a function of $N$ when $\vert\alpha\vert^2=\vert\beta\vert^2=1/2$ where $M_\star$ and $K_\star$ are the optimal values of $M$ and $K$ that maximize the quantum capacity $Q$ or equivalently the transfer efficiency $\zeta_\mathrm{q}$ for given $N$ as in Fig.~\ref{fig:Fidelity}.  The 50\,\% efficiency ($Q=1$ qubit/electron-photon) is attained when $N=218$ with $M_\star=21$ and $K_\star=15$.
    }
    \label{fig:QDC-QC-CAP}
\end{figure}

%
%
%

\section{Discussion}\label{sec 5}
The quantum duplex coding and telexchanging are full-duplex counterparts to the quantum superdense coding and teleportation, respectively. 
The quantum superdense coding utilizes a striking nonclassical property of Bell states, which sends two bits of classical information in one qubit whereas quantum teleportation enables remote parties to transfer quantum information with two-bit classical information by using the preshared quantum entanglement.
The quantum duplex coding and telexchanging have the following advantages over these quantum communication protocols. 

\begin{itemize}
\item 
Quantum superdense coding is a simplex protocol to allow the communication of a two-bit classical message in one direction only. In contrast, quantum duplex coding allows the full-duplex communication by means of the nonlocal \ac{DNOT} operation. Although both the protocols transmit 2 bits/Bell-pair, the main advantage of the quantum duplex coding is to transfer classical information in both directions simultaneously.
%
In quantum superdense coding, a sender needs to transmit a qubit over the quantum channel to enable a receiver to decode two bits of classical information.  In contrast, the quantum duplex coding enables remote parties to communicate without transmitting any physical particle over the channel at the cost of the erasure probability $1-\zeta_{\text{c}}$. As $N$ and $K$ increase, the capacity of the quantum duplex coding approaches to 2 bits/Bell-pair same as the capacity of quantum superdense coding. 
\begin{figure}[t!]
 \centering{
\includegraphics[width=0.55\textwidth]{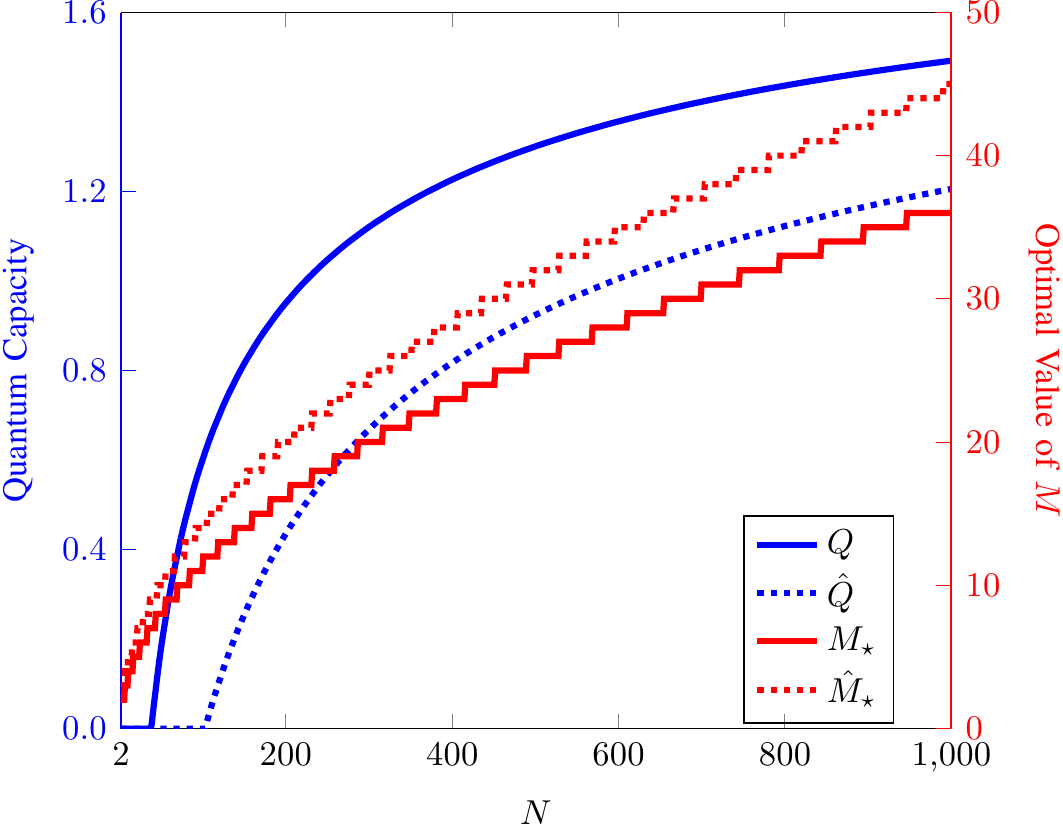}}
    \caption{
Quantum capacity and the optimal value of $M$ for the quantum telexchaning and the counterfactual state exchange protocol presented in \cite{LAYZ:19:PRA} where $K$ is simply set to $M_\star$ for the \ac{MQZ}-\ac{CQZ} telexchanging protocol. Here $\hat{Q}$ and $\hat{M}_\star$ denote the quantum capacity and the optimal value of $M$ for the protocol presented in \cite{LAYZ:19:PRA}. 
    }
    \label{fig:comparison}
\end{figure}

\item 
Quantum teleportation utilizes a prior entanglement to send one qubit of a quantum message. In contrast, quantum telexchanging enables communicating parties to transmit quantum information without preshared entanglement. To exchange a qubit of quantum information by using quantum teleportation, remote parties need two Bell pairs at the cost of four qubits and four bits of classical information.
In contrast, the quantum telexchanging enables remote parties to exchange one qubit quantum information at the cost of one ancilla qubit and one bit of classical information. To ensure the conterfactuality of the protocol, the quantum capacity of quantum telexchanging is limited to $Q$ in \eqref{eq: quantum capacity} due to the erasure probability $1-\zeta_{\text{q}}$. 
\end{itemize}
Recently, the counterfactual quantum state exchange has been demonstrated by using the \ac{CQZ} gates along with time-bin device \cite{LAYZ:19:PRA}. Fig.~\ref{fig:comparison} shows the comparison between quantum telexchanging (our scheme) and the protocol proposed in \cite{LAYZ:19:PRA} in terms of quantum capacity and optimal value of $M$ as a function of $N$ when $\vert\alpha\vert^2=\vert\gamma\vert^2=1/2$. Here $Q$ is the quantum capacity achieved by \ac{MQZ}-\ac{CQZ} telexchaning with $M=K=M_\star$ and $\hat{Q}$ is the quantum capacity achieved by the protocol in \cite{LAYZ:19:PRA} with $M=\hat{M}_\star$. It can be clearly seen that $Q>\hat{Q}$ whereas  $M_{\star}<\hat{M}_{\star}$ for all values of $N$, showing that our protocol gives better performance while using less resources (number of outer cycles).

\section{Conclusion}\label{sec 6}

We have put forth the new quantum communication protocols that achieve both full duplexity and counterfactuality for the classical as well as quantum information. Using the preshared entanglement and the nonlocal \ac{DNOT} operation (counterfactual disentanglement), this unique quantum protocol allows remote parties to swap a one-bit of classical information simultaneously without transmitting any physical particle over the channel. We have generalized this counterfactual duplex communication framework for the quantum information by devising  the dual \ac{DNOT} operation (counterfactual entanglement followed by disentanglement in a distributed way) along with local operations and one-bit  announcement of classical information.  The communication without transmitting any physical particle over the channel can further provide inherent security advantages over the most of eavesdropping attacks such as the photon-number splitting attack and the intercept-and-resend attack. The future work can be done to extend the quantum communication protocols to transfer the information in both directions simultaneously, counterfactually---and \emph{securely}.

\section*{Acknowledgment}

This work was supported in part by the the National Research Foundation of Korea under Grant  2019R1A2C2007037, ITRC (Information Technology Research Center) support program (IITP-2021-0-02046) supervised by the IITP (Institute for Information \& Communications Technology Planning \& Evaluation) and in part by the Office of Naval Research under Grant  N00014-19-1-2724.


\begin{thebibliography}{42}
\providecommand{\natexlab}[1]{#1}
\providecommand{\url}[1]{\texttt{#1}}
\expandafter\ifx\csname urlstyle\endcsname\relax
  \providecommand{\doi}[1]{doi: #1}\else
  \providecommand{\doi}{doi: \begingroup \urlstyle{rm}\Url}\fi

\bibitem[Zhang et~al.(2016)Zhang, Long, Vasilakos, and Hanzo]{ZLVH:16:IEEE}
Zhongshan Zhang, Keping Long, Athanasios~V Vasilakos, and Lajos Hanzo.
\newblock Full-duplex wireless communications: Challenges, solutions, and
  future research directions.
\newblock \emph{Proc. IEEE}, 104\penalty0 (7):\penalty0 1369--1409, July 2016.
\newblock \doi{10.1109/JPROC.2015.2497203}.

\bibitem[Wang et~al.(2016)Wang, Zhang, Liu, and Hu]{WZLH:16:QIP}
He~Wang, Yu~Qing Zhang, Xue~Feng Liu, and Yu~Pu Hu.
\newblock Efficient quantum dialogue using entangled states and entanglement
  swapping without information leakage.
\newblock \emph{Quantum Inf Process}, 15\penalty0 (6):\penalty0 2593--2603,
  March 2016.
\newblock \doi{10.1007/s11128-016-1294-z}.

\bibitem[Maitra(2017)]{M:17:QIP}
Arpita Maitra.
\newblock Measurement device-independent quantum dialogue.
\newblock \emph{Quantum Inf Process}, 16\penalty0 (12):\penalty0 1--15,
  November 2017.
\newblock \doi{10.1007/s11128-017-1757-x}.

\bibitem[Zhou et~al.(2020)Zhou, Sheng, and Long]{ZSYL:20:SB}
Lan Zhou, Yu-Bo Sheng, and Gui-Lu Long.
\newblock Device-independent quantum secure direct communication against
  collective attacks.
\newblock \emph{Sci. Bull.}, 65\penalty0 (1):\penalty0 12--20, January 2020.
\newblock \doi{10.1016/j.scib.2019.10.025}.

\bibitem[Zhang et~al.(2017)Zhang, Ding, Sheng, Zhou, Shi, and
  Guo]{ZDSZSG:17:PRL}
Wei Zhang, Dong-Sheng Ding, Yu-Bo Sheng, Lan Zhou, Bao-Sen Shi, and Guang-Can
  Guo.
\newblock Quantum secure direct communication with quantum memory.
\newblock \emph{Phys. Rev. Lett.}, 118\penalty0 (22):\penalty0 220501, May
  2017.
\newblock \doi{10.1103/PhysRevLett.118.220501}.

\bibitem[Lee et~al.(2019{\natexlab{a}})Lee, Takagi, Yamasaki, Adesso, and
  Lee]{LTYAL:19:PRL}
Yonghae Lee, Ryuji Takagi, Hayata Yamasaki, Gerardo Adesso, and Soojoon Lee.
\newblock State exchange with quantum side information.
\newblock \emph{Phys. Rev. Lett.}, 122:\penalty0 010502, January
  2019{\natexlab{a}}.
\newblock \doi{10.1103/PhysRevLett.122.010502}.

\bibitem[Lee et~al.(2019{\natexlab{b}})Lee, Yamasaki, Adesso, and
  Lee]{LYAL:19:PRA}
Yonghae Lee, Hayata Yamasaki, Gerardo Adesso, and Soojoon Lee.
\newblock One-shot quantum state exchange.
\newblock \emph{Phys. Rev. A}, 100:\penalty0 042306, October
  2019{\natexlab{b}}.
\newblock \doi{10.1103/PhysRevA.100.042306}.

\bibitem[Oppenheim and Winter(2005)]{OW:05:arXiv}
Jonathan Oppenheim and Andreas Winter.
\newblock Uncommon information (the cost of exchanging a quantum state).
\newblock \emph{arXiv:0511082}, 2005.

\bibitem[Li et~al.(2019)Li, Al-Amri, Yang, and Zubairy]{LAYZ:19:PRA}
Zheng-Hong Li, M.~Al-Amri, Xi-Hua Yang, and M.~Suhail Zubairy.
\newblock Counterfactual exchange of unknown quantum states.
\newblock \emph{Phys. Rev. A}, 100:\penalty0 022110, August 2019.
\newblock \doi{10.1103/PhysRevA.100.022110}.

\bibitem[Farooq et~al.(2019)Farooq, ur~Rehman, Jeong, Kim, and
  Shin]{FRJKS:19:SR}
Ahmad Farooq, Junaid ur~Rehman, Youngmin Jeong, Jeong~San Kim, and Hyundong
  Shin.
\newblock Tightening monogamy and polygamy inequalities of multiqubit
  entanglement.
\newblock \emph{Sci. Rep.}, 9:\penalty0 3314, March 2019.
\newblock \doi{10.1038/s41598-018-37731-z}.

\bibitem[Horodecki et~al.(2009)Horodecki, Horodecki, Horodecki, and
  Horodecki]{HHHH:09:RMP}
R.~Horodecki, P.~Horodecki, M.~Horodecki, and K.~Horodecki.
\newblock Quantum entanglement.
\newblock \emph{Rev. Mod. Phys.}, 81\penalty0 (2):\penalty0 865--942, June
  2009.
\newblock \doi{10.1103/RevModPhys.81.865}.

\bibitem[Barreiro et~al.(2005)Barreiro, Langford, Peters, and
  Kwiat]{BLPK:05:PRL}
Julio~T Barreiro, Nathan~K Langford, Nicholas~A Peters, and Paul~G Kwiat.
\newblock Generation of hyperentangled photon pairs.
\newblock \emph{Phys. Rev. Lett.}, 95\penalty0 (26):\penalty0 260501, December
  2005.
\newblock \doi{10.1103/PhysRevLett.95.260501}.

\bibitem[Salih et~al.(2013)Salih, Li, Al-Amri, and Zubairy]{SLAZ:13:PRL}
Hatim Salih, Zheng-Hong Li, M~Al-Amri, and M~Suhail Zubairy.
\newblock Protocol for direct counterfactual quantum communication.
\newblock \emph{Phys. Rev. Lett.}, 110\penalty0 (17):\penalty0 170502, April
  2013.
\newblock \doi{10.1103/PhysRevLett.110.170502}.

\bibitem[Aharonov and Vaidman(2019)]{AV:19:PRA}
Yakir Aharonov and Lev Vaidman.
\newblock Modification of counterfactual communication protocols that
  eliminates weak particle traces.
\newblock \emph{Phys. Rev. A}, 99\penalty0 (1):\penalty0 010103, January 2019.
\newblock \doi{10.1103/PhysRevA.99.010103}.

\bibitem[Li et~al.(2014)Li, Al-Amri, and Zubairy]{LAZ:14:PRA}
Zheng-Hong Li, M.~Al-Amri, and M.~Suhail Zubairy.
\newblock Direct quantum communication with almost invisible photons.
\newblock \emph{Phys. Rev. A}, 89:\penalty0 052334, May 2014.
\newblock \doi{10.1103/PhysRevA.89.052334}.

\bibitem[Cao et~al.(2017)Cao, Li, Cao, Yin, Chen, Yin, Chen, Ma, Peng, and
  Pan]{CLYCYCMPP:17:PNAS}
Yuan Cao, Yu-Huai Li, Zhu Cao, Juan Yin, Yu-Ao Chen, Hua-Lei Yin, Teng-Yun
  Chen, Xiongfeng Ma, Cheng-Zhi Peng, and Jian-Wei Pan.
\newblock Direct counterfactual communication via quantum {Zeno} effect.
\newblock \emph{Proc Natl Acad Sci USA}, 114\penalty0 (19):\penalty0
  4920--4924, May 2017.
\newblock \doi{10.1073/pnas.1614560114}.

\bibitem[Aharonov et~al.(2021)Aharonov, Cohen, and Popescu]{ACP:21:NC}
Yakir Aharonov, Eliahu Cohen, and Sandu Popescu.
\newblock A dynamical quantum cheshire cat effect and implications for
  counterfactual communication.
\newblock \emph{Nature Communications}, 12\penalty0 (1):\penalty0 1--8, August
  2021.
\newblock \doi{10.1038/s41467-021-24933-9}.

\bibitem[Hosten et~al.(2006)Hosten, Rakher, Barreiro, Peters, and
  Kwiat]{HRBPK:06:Nature}
Onur Hosten, Matthew~T Rakher, Julio~T Barreiro, Nicholas~A Peters, and Paul~G
  Kwiat.
\newblock Counterfactual quantum computation through quantum interrogation.
\newblock \emph{Nature}, 439\penalty0 (7079):\penalty0 949, February 2006.
\newblock \doi{10.1038/nature04523}.

\bibitem[Kong et~al.(2015)Kong, Ju, Huang, Wang, Kong, Shi, Jiang, and
  Du]{KJHWKSJD:15:PRL}
Fei Kong, Chenyong Ju, Pu~Huang, Pengfei Wang, Xi~Kong, Fazhan Shi, Liang
  Jiang, and Jiangfeng Du.
\newblock Experimental realization of high-efficiency counterfactual
  computation.
\newblock \emph{Phys. Rev. Lett.}, 115\penalty0 (8):\penalty0 080501, August
  2015.
\newblock \doi{10.1103/PhysRevLett.115.080501}.

\bibitem[Mitchison et~al.(2007)Mitchison, Jozsa, and Popescu]{MJP:07:PRA}
Graeme Mitchison, Richard Jozsa, and Sandu Popescu.
\newblock Sequential weak measurement.
\newblock \emph{Phys. Rev. A}, 76:\penalty0 062105, December 2007.
\newblock \doi{10.1103/PhysRevA.76.062105}.

\bibitem[Zaman et~al.(2020)Zaman, Shin, and Win]{ZSW:20:arXiv}
Fakhar Zaman, Hyundong Shin, and Moe~Z. Win.
\newblock Counterfactual concealed telecomputation.
\newblock \emph{arXiv:2012.04948}, 2020.

\bibitem[Noh(2009)]{N:09:PRL}
Tae-Gon Noh.
\newblock Counterfactual quantum cryptography.
\newblock \emph{Phys. Rev. Lett.}, 103\penalty0 (23):\penalty0 230501, December
  2009.
\newblock \doi{10.1103/PhysRevLett.103.230501}.

\bibitem[Yin et~al.(2012)Yin, Li, Yao, Zhang, Wang, Chen, Guo, and
  Han]{YLYZWCGH:12:PRA}
Zhen-Qiang Yin, Hong-Wei Li, Yao Yao, Chun-Mei Zhang, Shuang Wang, Wei Chen,
  Guang-Can Guo, and Zheng-Fu Han.
\newblock Counterfactual quantum cryptography based on weak coherent states.
\newblock \emph{Phys. Rev. A}, 86\penalty0 (2):\penalty0 022313, August 2012.
\newblock \doi{10.1103/PhysRevA.86.022313}.

\bibitem[Salih(2014)]{SH:14:PRA}
Hatim Salih.
\newblock Tripartite counterfactual quantum cryptography.
\newblock \emph{Phys. Rev. A}, 90\penalty0 (1):\penalty0 012333, July 2014.
\newblock \doi{10.1103/PhysRevA.90.012333}.

\bibitem[Sun and Wen(2010)]{SW:10:PRA}
Ying Sun and Qiao-Yan Wen.
\newblock Counterfactual quantum key distribution with high efficiency.
\newblock \emph{Phys. Rev. A}, 82\penalty0 (5):\penalty0 052318, November 2010.
\newblock \doi{10.1103/PhysRevA.82.052318}.

\bibitem[Liu et~al.(2012)Liu, Ju, Liang, Tang, Tu, Zhou, Peng, Chen, Chen,
  Chen, and Pan]{LJLTTZPCCCP:12:PRL}
Yang Liu, Lei Ju, Xiao-Lei Liang, Shi-Biao Tang, Guo-Liang~Shen Tu, Lei Zhou,
  Cheng-Zhi Peng, Kai Chen, Teng-Yun Chen, Zeng-Bing Chen, and Jian-Wei Pan.
\newblock Experimental demonstration of counterfactual quantum communication.
\newblock \emph{Phys. Rev. Lett.}, 109:\penalty0 030501, July 2012.
\newblock \doi{10.1103/PhysRevLett.109.030501}.

\bibitem[Elitzur and Vaidman(1993)]{VE:93:FOP}
Avshalom Elitzur and Lev Vaidman.
\newblock Quantum mechanical interaction-free measurement.
\newblock \emph{Found. Phys.}, 23\penalty0 (76):\penalty0 987--997, July 1993.
\newblock \doi{10.1007/BF00736012}.

\bibitem[Kwiat et~al.(1995)Kwiat, Weinfurter, Herzog, Zeilinger, and
  Kasevich]{KWHZK:95:PRL}
Paul Kwiat, Harald Weinfurter, Thomas Herzog, Anton Zeilinger, and Mark~A
  Kasevich.
\newblock Interaction-free measurement.
\newblock \emph{Phys. Rev. Lett.}, 74\penalty0 (24):\penalty0 4763, November
  1995.
\newblock \doi{10.1103/PhysRevLett.74.4763}.

\bibitem[Calafell et~al.(2019)Calafell, Str{\"o}mberg, Arvidsson-Shukur,
  Rozema, Saggio, Greganti, Harris, Prabhu, Carolan, Hochberg,
  et~al.]{Cetal:19:NPJ}
I~Alonso Calafell, T~Str{\"o}mberg, DRM Arvidsson-Shukur, LA~Rozema, V~Saggio,
  C~Greganti, NC~Harris, M~Prabhu, J~Carolan, M~Hochberg, et~al.
\newblock Trace-free counterfactual communication with a nanophotonic
  processor.
\newblock \emph{npj Quantum Information}, 5\penalty0 (1):\penalty0 1--5, 2019.
\newblock \doi{10.1038/s41534-019-0179-2}.

\bibitem[Guo et~al.(2015)Guo, Cheng, Chen, Wang, and Zhang]{GCCWHZ:15:SR}
Qi~Guo, Liu-Yong Cheng, Li~Chen, Hong-Fu Wang, and Shou Zhang.
\newblock Counterfactual quantum-information transfer without transmitting any
  physical particles.
\newblock \emph{Sci. Rep.}, 5:\penalty0 8416, February 2015.
\newblock \doi{10.1038/srep08416}.

\bibitem[Li et~al.(2015)Li, Al-Amri, and Zubairy]{LAZ:15:PRA}
Zheng-Hong Li, M~Al-Amri, and M~Suhail Zubairy.
\newblock Direct counterfactual transmission of a quantum state.
\newblock \emph{Phys. Rev. A}, 92\penalty0 (5):\penalty0 052315, November 2015.
\newblock \doi{10.1103/PhysRevA.92.052315}.

\bibitem[Salih(2016)]{SH:16:FP}
Hatim Salih.
\newblock Protocol for counterfactually transporting an unknown qubit.
\newblock \emph{Front. Phys.}, 3:\penalty0 94, January 2016.
\newblock \doi{10.3389/fphy.2015.00094}.

\bibitem[Guo et~al.(2014)Guo, Cheng, Chen, Wang, and Zhang]{GCCWZ:14:OPT}
Qi~Guo, Liu-Yong Cheng, Li~Chen, Hong-Fu Wang, and Shou Zhang.
\newblock Counterfactual entanglement distribution without transmitting any
  particles.
\newblock \emph{Opt. Express}, 22\penalty0 (8):\penalty0 8970--8984, April
  2014.
\newblock \doi{10.1364/OE.22.008970}.

\bibitem[Chen et~al.(2015)Chen, Gu, Jiang, Xie, and Chen]{CGJXC:15:OPT}
Yuanyuan Chen, Xuemei Gu, Dong Jiang, Ling Xie, and Lijun Chen.
\newblock Tripartite counterfactual entanglement distribution.
\newblock \emph{Opt. Express}, 23\penalty0 (16):\penalty0 21193--21203, August
  2015.
\newblock \doi{10.1364/OE.23.021193}.

\bibitem[Chen et~al.(2016)Chen, Jian, Gu, Xie, and Chen]{CJGXC:16:JOSA}
Yuanyuan Chen, Dong Jian, Xuemei Gu, Ling Xie, and Lijun Chen.
\newblock Counterfactual entanglement distribution using quantum dot spins.
\newblock \emph{JOSA B}, 33\penalty0 (4):\penalty0 663--669, January 2016.
\newblock \doi{10.1364/JOSAB.33.000663}.

\bibitem[Zaman et~al.(2018)Zaman, Jeong, and Shin]{ZJS:18:SR}
Fakhar Zaman, Youngmin Jeong, and Hyundong Shin.
\newblock Counterfactual {Bell}-state analysis.
\newblock \emph{Sci. Rep.}, 8\penalty0 (1):\penalty0 14641, October 2018.
\newblock \doi{10.1038/s41598-018-32928-8}.

\bibitem[Zaman et~al.(2021)Zaman, Hong, and Shin]{ZHS:21:QIP}
Fakhar Zaman, Een-Kee Hong, and Hyundong Shin.
\newblock Local distinguishability of {Bell}-type states.
\newblock \emph{Quantum Inf Process}, 20\penalty0 (5):\penalty0 1--12, May
  2021.
\newblock \doi{10.1007/s11128-021-03114-z}.

\bibitem[Guo et~al.(2017)Guo, Zhai, Cheng, Wang, and Zhang]{HZCWZ:17:PRA}
Qi~Guo, Shuqin Zhai, Liu-Yong Cheng, Hong-Fu Wang, and Shou Zhang.
\newblock Counterfactual quantum cloning without transmitting any physical
  particles.
\newblock \emph{Phys. Rev. A}, 96:\penalty0 052335, November 2017.
\newblock \doi{10.1103/PhysRevA.96.052335}.

\bibitem[Itano et~al.(1990)Itano, Heinzen, Bollinger, and
  Wineland]{IHBW:90:PRA}
Wayne~M Itano, Daniel~J Heinzen, J~J Bollinger, and DJ~Wineland.
\newblock Quantum {Zeno} effect.
\newblock \emph{Phys. Rev. A}, 41\penalty0 (5):\penalty0 2295, March 1990.
\newblock \doi{10.1103/PhysRevA.41.2295}.

\bibitem[Zaman et~al.(2019)Zaman, Jeong, and Shin]{ZJS:19:SR}
Fakhar Zaman, Youngmin Jeong, and Hyundong Shin.
\newblock Dual quantum {Zeno} superdense coding.
\newblock \emph{Sci. Rep.}, 9\penalty0 (1):\penalty0 11193, August 2019.
\newblock \doi{10.1038/s41598-019-47667-7}.

\bibitem[Cover and Thomas(2012)]{CT:12:JW}
Thomas~M Cover and Joy~A Thomas.
\newblock \emph{Elements of {Information} {Theory}}.
\newblock Wiley, New York, NY, USA, 2012.
\newblock \doi{10.1002/047174882X}.

\bibitem[Bennett et~al.(1997)Bennett, DiVincenzo, and Smolin]{BDS:97:PRL}
Charles~H Bennett, David~P DiVincenzo, and John~A Smolin.
\newblock Capacities of quantum erasure channels.
\newblock \emph{Phys. Rev. Lett.}, 78\penalty0 (16):\penalty0 3217, April 1997.
\newblock \doi{10.1103/PhysRevLett.78.3217}.

\end{thebibliography}
\end{document}